\documentclass{IEEEtran}
\usepackage{latexsym}
\usepackage{graphicx}
\usepackage{amsfonts,amssymb,amsmath}
\usepackage{hyperref}
\usepackage{cite}
\usepackage{url}
\usepackage{multirow}
\usepackage{algorithmic}
\usepackage{ulem}
\usepackage{balance}
\usepackage{textcomp}
\usepackage{scalerel}
\usepackage{color}
\usepackage{subfigure}
\def\BibTeX{{\rm B\kern-.05em{\sc i\kern-.025em b}\kern-.08em
    T\kern-.1667em\lower.7ex\hbox{E}\kern-.125emX}}
\begin{document}
\title{Environment-Aware Channel Measurement and Modeling for Terahertz Monostatic Sensing}
\author{Yejian Lyu, Zhiqiang Yuan, Henk Wymeersch,~\textit{Fellow, IEEE}, and Chong Han,~\textit{Senior Member, IEEE} 
\thanks{

Y. Lyu and C. Han are with the Terahertz Wireless Communications Laboratory at Shanghai Jiao Tong University, Shanghai 200240, China (e-mail: \{yejian.lyu; chong.han\}@sjtu.edu.cn);

Z. Yuan and H. Wymeersch are with the Communication Systems Group at Chalmers University of Technology, Gothenburg, Sweden.
}}
\maketitle
\begin{abstract}
Integrated sensing and communication (ISAC) at terahertz (THz) frequencies holds significant promise for unifying ultra-high-speed wireless connectivity with fine-grained environmental awareness. Realistic and interpretable channel modeling is essential to fully realize the potential of such systems. This work presents a comprehensive investigation of monostatic sensing channels at 300~GHz, based on an extensive measurement campaign conducted at 57 co-located transceiver (TRx) positions across three representative indoor scenarios. Multipath component (MPC) parameters, including amplitude, delay, and angle, are extracted using a high-resolution space-alternating generalized expectation-maximization (SAGE) algorithm. To cluster the extracted MPCs, an image-processing-based clustering method, i.e., connected component labeling (CCL), is applied to group MPCs based on delay-angle consistency. Based on the measurement data, an environment-aware channel modeling framework is proposed to establish mappings between physical scenario attributes (e.g., reflector geometry, surface materials, and roughness) and their corresponding channel-domain manifestations. The framework incorporates both specular and diffuse reflections and leverages several channel parameters, e.g., reflection loss, Lambertian scattering, and intra-cluster dispersion models, to characterize reflection behavior. Experimental results demonstrate that the proposed approach can reliably extract physical characteristics, e.g., structural and material information, from the observed channel characteristics, offering a promising foundation for advanced THz ISAC channel modeling.
\end{abstract}


\section{Introduction}
\label{sec:introduction}
\IEEEPARstart{I}{ntegrated} sensing and communication (ISAC) is a key enabler for sixth-generation ($6$G) and beyond wireless systems, allowing networks to sense and interpret their physical environment~\cite{isac_jhzhang,6G_isac1,6G_isac2,isac_low_freq_model,intro}. The advancement of terahertz (THz) technology has extended ISAC capabilities to high-resolution tasks such as human activity recognition, automotive radar, and indoor mapping~\cite{THz_sensing1,THz_sensing2,THz_sensing3}. The THz band, spanning $100$~GHz to $10$~THz, offers ultra-wide bandwidth and sub-millimeter wavelengths. These characteristics, together with large antenna arrays or highly directional antennas, enable fine delay and angular resolution~\cite{dupleich2024characterization,wu2021thz,thz_isac_importance}. Compared to microwave and free-space optical systems, THz signals strike a favorable trade-off between spatial accuracy and propagation robustness, making them well-suited for radio-based environmental sensing in complex indoor scenarios. Notably, THz can identify material properties, which are not available in other frequency bands, due to its unique interaction with matter, exhibiting strong frequency-dependent absorption and reflection that reveal material-specific spectral fingerprints.

Accurate and reliable channel modeling serves as a critical foundation for designing and evaluating emerging wireless communication systems, especially those integrating sensing functionalities~\cite{channel_sounder_des,alma9920694098005762}. Wireless sensing scenarios are generally classified based on transmitter (Tx) and receiver (Rx) placement into two main configurations: monostatic, where Tx and Rx share the same location, and bistatic, where they occupy distinct spatial positions~\cite{sensing_case}. Compared with conventional bistatic communication channels, which exhibit complex multipath propagation due to separated Tx-Rx links, monostatic sensing channels primarily involve direct reflections, both specular and diffuse, from environmental objects~\cite{mono_sensor}. Due to these fundamental differences, it becomes critical to characterize monostatic sensing channel features in detail.

\begin{table*}[!t]
\caption{Summary of State-of-the-Art THz Sensing Channel Measurements and Modeling}\label{tab:related_work}
\label{tab:related_work}
\centering
\begin{tabular}{p{1.2cm} p{1.7cm} p{3cm} p{3cm} p{5.8cm}}
\hline
\textbf{Ref.} & \textbf{Frequency} & \textbf{Sensing Mode} & \textbf{Scenario} & \textbf{Analysis} \\
\hline
\cite{hybrid_isac} & 130-143~GHz & Monostatic \& bistatic & Indoor \& outdoor &  Scatterer-based hybrid ISAC model \\
\cite{dupleich2024characterization} & 190~GHz & Bistatic & Factory &  Scatterer detection based on power-angle spectrum \\
\cite{thz_mono_sen_ybli} & 306-321~GHz & Monostatic & Empty room \& corridors & Ranging accuracy and back-scattering coefficient \\
\cite{lotti2023radio} & 235-320~GHz & Monostatic & Laboratory & SLAM-based mapping and tracking \\
\cite{monostatic_lyu} & 290-310~GHz & Monostatic & Laboratory & THz hybrid sensing channel model \\
\cite{fang_geometry} & 290-310~GHz & Monostatic & Laboratory & Centimeter-level Environment reconstruction\\
This work & 290-310~GHz & Monostatic & Three distinct indoor scenarios & Cluster-based environment-aware channel models for THz sensing\\ 
\hline
\end{tabular}
\end{table*}

Despite growing interest in THz sensing, current literature addressing monostatic and bistatic channel measurements at THz frequencies remains limited. For instance, previous studies have investigated combined bistatic and monostatic channels at $140$~GHz, developing hybrid deterministic-stochastic models to describe propagation in both indoor and outdoor environments~\cite{hybrid_isac}. At $190$~GHz, bistatic sensing measurements were conducted in an industrial scenario, demonstrating the feasibility of high-resolution detection of environmental scatterers~\cite{dupleich2024characterization}. Additionally, indoor monostatic channel measurements at frequencies between $306$ and $321$~GHz have assessed ranging accuracy and back-scattering characteristics in various environments, including empty rooms and corridors~\cite{thz_mono_sen_ybli}. Besides, recent work at $235$-$320$~GHz utilized radio simultaneous localization and mapping frameworks to map indoor environments and track monostatic transceiver (TRx) positions~\cite{lotti2023radio}. In our previous work~\cite{monostatic_lyu}, monostatic sensing channel measurements were conducted in an indoor laboratory environment at frequencies ranging from $290$ to $310$~GHz. A hybrid modeling approach was introduced, where diffuse components were leveraged for environmental reconstruction and specular reflections were utilized for preliminary material identification. Building upon this, environmental reconstruction was further implemented and evaluated in~\cite{fang_geometry} using the same measurement dataset. However, the scope of material identification in~\cite{monostatic_lyu,fang_geometry} was limited, focusing only on distinguishing between cement and metal surfaces, without addressing a broader range of common indoor materials. A summary of the state-of-the-art works for THz sensing channels modeling are presented in Table~\ref{tab:related_work}.

Despite considerable progress above, several critical aspects remain insufficiently addressed in the current literature. First, despite significant progress in THz bistatic channel research, studies on THz monostatic sensing channels remain relatively scarce in the existing literature. Second, detailed identification and analysis of cluster-based channel characteristics, such as intra-cluster delay and angular spreads, and power and reflection loss distributions, remain underexplored in the context of THz sensing channels, limiting the accuracy of existing channel models. Third, although several studies have investigated the material properties using THz frequencies~\cite{thz_material1,thz_material2,material1,material2,fang_geometry}, existing studies rarely establish explicit links between physical environmental features, such as object geometry, surface roughness, and material composition, and their corresponding channel characteristics in the THz frequency range. To fill the aforementioned gaps, this work investigates monostatic sensing channels at $300$~GHz by focusing explicitly on material-specific characteristics in three typical indoor scenarios. Through extensive measurement campaigns and a novel environment-aware modeling framework, we bridge the gap between physical material characteristics and their channel-domain representations, providing unique insights and models tailored specifically for THz monostatic sensing. The key contributions of this paper are summarized as follows:

\begin{itemize}
\item A comprehensive THz monostatic sensing measurement dataset is presented at the $300$~GHz band, collected across three structurally and materially diverse indoor environments with $57$ co-located TRx positions. Unlike prior datasets that are limited to simplified or homogeneous scenarios, the proposed dataset captures a rich variety of spatial, geometric, and material characteristics, including metallic, dielectric, and composite surfaces. This diversity enables high-fidelity modeling and simulation of realistic monostatic sensing configurations. The dataset provides a valuable foundation for advancing research in channel modeling, material-aware sensing, localization, and ISAC, supporting the development and evaluation of novel algorithms for future THz-enabled applications.

\item This work advances the analysis of THz monostatic sensing channels by providing a detailed statistical characterization at the multipath component (MPC) cluster level, going beyond the TRx-location-specific metrics (e.g., delay and angular spreads at individual TRx positions) reported in~\cite{monostatic_lyu}. Using a high-resolution parameter estimation algorithm (SAGE), multipath components are extracted and grouped to analyze cluster-wise delay and angular characteristics. Based on this enriched dataset, we propose a statistical channel model for indoor monostatic sensing, which captures not only global channel parameters but also intra-cluster properties. This modeling approach enables more realistic evaluations of sensing performance, which are critical for ISAC system functions, e.g., object detection, localization, and environment reconstruction.

\item Building on the statistical cluster-based channel characterization, we develop an environment-aware channel modeling framework for monostatic sensing, which explicitly links physical characteristics of the environment, such as material type, surface roughness, object geometry, and scene complexity, to corresponding propagation features. 
\end{itemize}
\begin{figure*}
    \centering
    \subfigure []{\includegraphics[width=0.9\columnwidth]{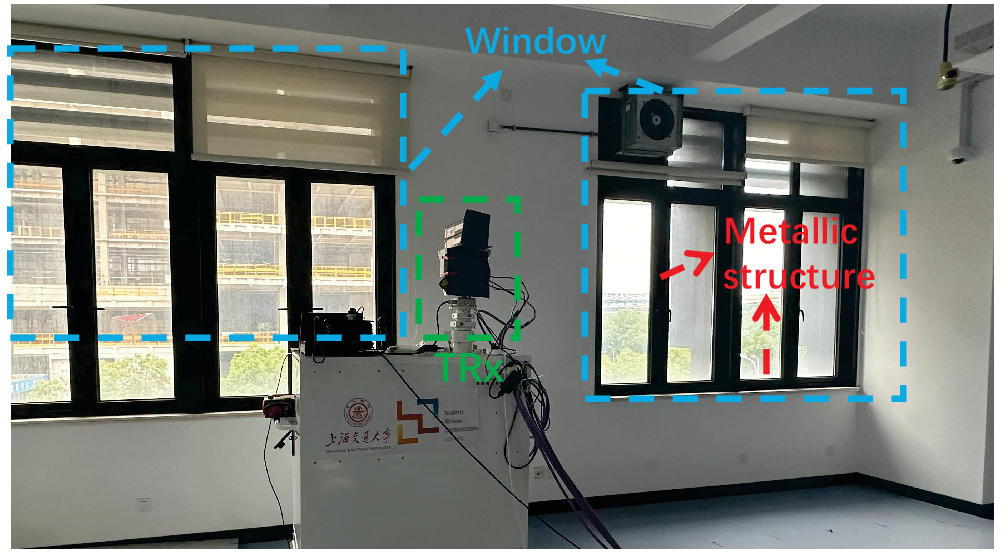}} \hspace{1.5cm}
    \subfigure []{\includegraphics[width=0.5\columnwidth]{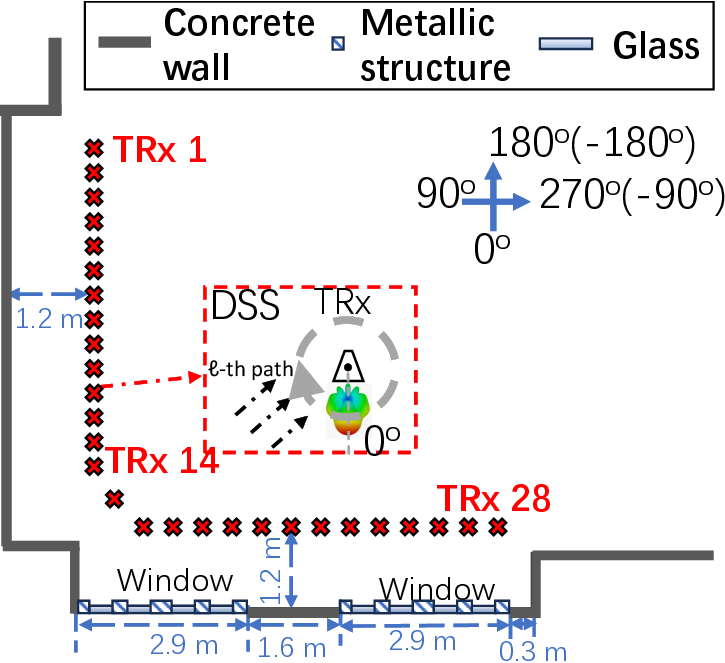}}
    \subfigure []{\includegraphics[width=0.65\columnwidth]{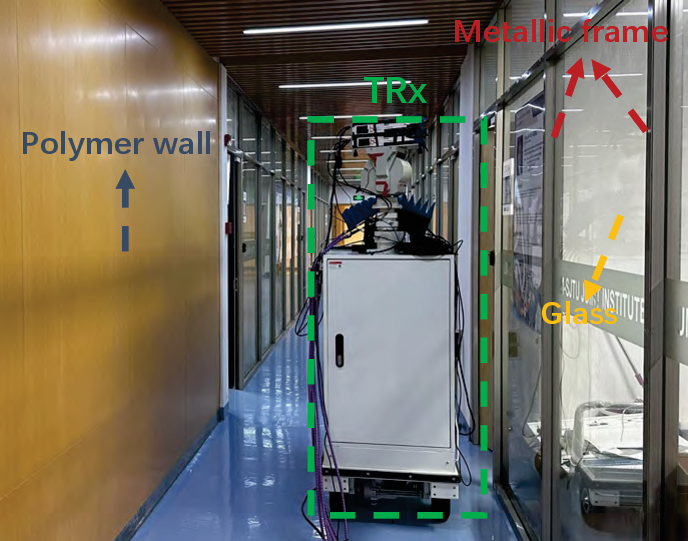}} \hspace{0.5cm}
    \subfigure []{\includegraphics[width=0.9\columnwidth]{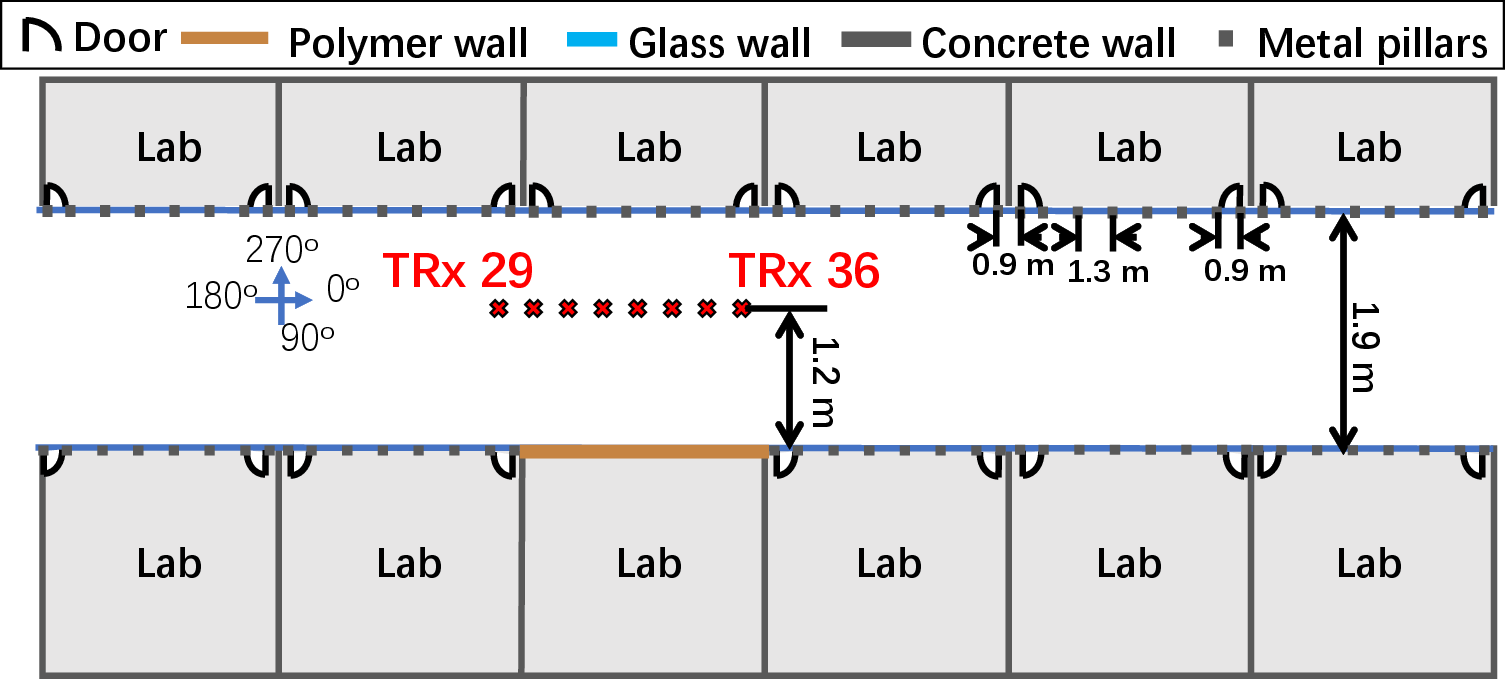}}
    \subfigure []{\includegraphics[width=0.83\columnwidth]{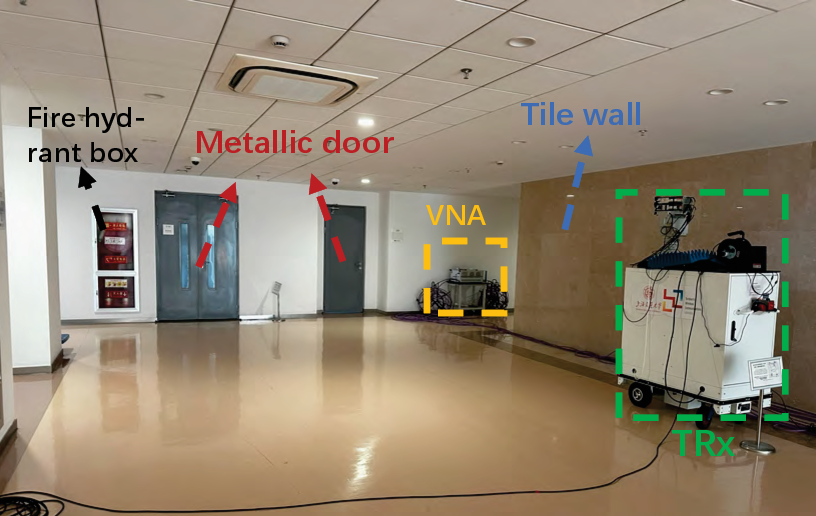}}\hspace{1cm}
    \subfigure []{\includegraphics[width=0.75\columnwidth]{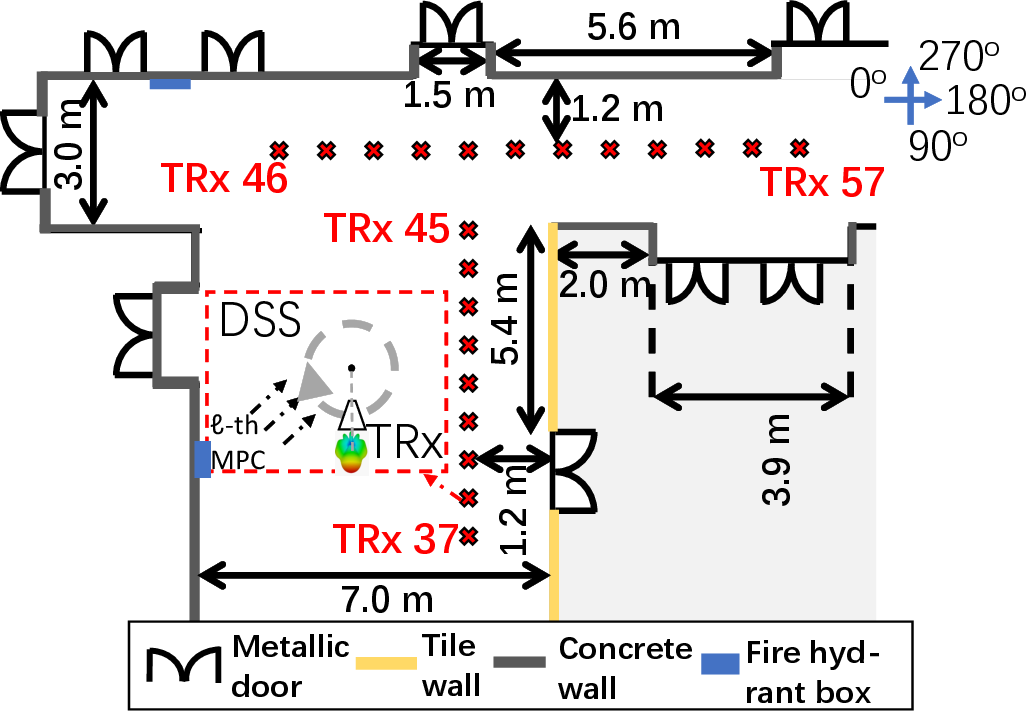}}
\caption {Diagrams and photos of the measurement scenario for THz monostatic sensing. (a) Real photo in Scenario $1$. (b) Diagram of Scenario $1$. (c) Real photo in Scenario $2$. (d) Diagram of Scenario $2$. (e) Real photo in Scenario $3$. (f) Diagram of Scenario $3$.}\label{fig:meas_pic}
\end{figure*}

The remainder of this paper is structured as follows. Section~\ref{sec:sounder} outlines the monostatic-sensing-based measurement setup and scenario. The data processing and image-processing-assisted cluster identification method are described in Section~\ref{sec:estimator}. The indoor channel characterization for THz monostatic sensing is presented in Section~\ref{sec:modeling}. Besides, a material-related modeling framework is proposed in in Section~\ref{sec:material}. Finally, concluding remarks are presented in Section~\ref{sec:conclusion}.

\section{Measurement Campaign}\label{sec:sounder}
In this section, we present the details of the THz monostatic channel measurement campaign. We begin by describing the measurement system setup, including the system configuration, antenna specifications, and calibration procedures. Subsequently, we introduce the three representative indoor scenarios used in the campaign, highlighting their spatial dimensions, layout structures, and dominant material compositions.
\subsection{Sounder and Measurement Setup}
In this work, a vector network analyzer (VNA)-based channel sounder~\cite{lyu_vna,yuanbo_sage} operating from $290$ to $310$~GHz is employed for monostatic sensing channel measurements. The system provides an ultra-wide bandwidth of $20$~GHz, corresponding to a fine delay resolution of approximately $1.5$~cm in the delay domain. A total of $2001$ frequency points are sampled, supporting a maximum detectable range of $30$~m. To ensure a high dynamic range and measurement accuracy, the intermediate frequency (IF) bandwidth is configured at $1$~kHz.

As illustrated in Fig.~\ref{fig:meas_pic}, the co-located Tx and Rx are mounted on a precision-controlled turntable to mimic monostatic sensing conditions. Identical high-gain horn antennas are used at both ends, featuring $25.5$~dBi gain and a narrow half-power beamwidth (HPBW) of $8^{\circ}$, enabling high angular resolution. The directional scanning sounding (DSS) scheme is applied to capture rich azimuthal characteristics, with antenna rotation covering $[0^{\circ}, 359^{\circ}]$ in fine $1^{\circ}$ increments, offering dense angular sampling. The antenna phase center is positioned $0.2$~m from the rotation axis to avoid mechanical shadowing and maintain spatial consistency. The Tx and Rx are set at heights of $2.0$~m and $2.1$~m, respectively. Prior to measurements, full system calibration is performed using a back-to-back setup to remove system response and ensure accurate channel estimation. A summary of key measurement parameters is provided in Table~\ref{tab:configuration}.

\begin{table}
\centering
\caption{The monostatic sensing channel measurement configuration for the three scenarios.}\label{tab:configuration}
\begin{tabular}{c|c}
\hline
Parameter   &  Value\\
\hline
\hline
Frequency range & $290$-$310$~GHz\\
\hline
Frequency bandwidth      & $20$~GHz\\
\hline
Frequency point      & $2001$\\
\hline
Maximum detectable distance &  $30$~m\\
\hline
IF bandwidth            & $1$~kHz\\
\hline
Tx \& Rx antenna type   & Horn\\
\hline
Antenna gain   & $25.5$~dBi\\
\hline
Azimuthal HPBW           & $8^{\circ}$\\
\hline
Rotation step     & $1^{\circ}$\\
\hline
Azimuthal rotation range  & $[0^{\circ}:359^{\circ}]$\\
\hline
Tx Antenna height     & $2.0$~m\\
\hline
Rx Antenna height     & $2.1$~m\\
\hline
\end{tabular}
\end{table}

\subsection{Measurement Scenario}
In this work, comprehensive monostatic channel measurements are conducted across three representative indoor scenarios to characterize diverse propagation conditions relevant to THz ISAC systems. Scenario~$1$ represents an indoor laboratory environment with an L-shaped measurement trajectory, as illustrated in Fig.~\ref{fig:meas_pic}~(a) and (b). Along this route, a total of $28$ TRx positions are uniformly spaced at intervals of $0.5$~m. As highlighted in Fig.~\ref{fig:meas_pic}~(b), the primary reflective objects in Scenario~$1$ include a flat cement wall and two metal-framed windows. The readers can refer to our previous work~\cite{monostatic_lyu} for a detailed description of this scenario.

Scenario~$2$ captures a narrow corridor scenario with a width of $1.9$~m. One side of the corridor is constructed with large glass panels supported by multiple small metal pillars, while the opposite side is a polymer wall. As depicted in Fig.~\ref{fig:meas_pic}~(c) and (d), eight TRx locations are placed along the corridor at $1$~m intervals and a fixed distance of $1.2$~m to the polymer wall, capturing the wave interactions with the polymer, transparent, and metallic materials under quasi-parallel wall alignment.

Scenario~$3$ represents an L-shaped hallway with a more complex architectural layout. The scenario features a mix of surface types, including concrete walls, ceramic tile finishes, fire hydrant boxes, and metallic doors, offering diverse reflection characteristics, as shown in Fig.~\ref{fig:meas_pic}~(e) and (f). In this scenario, $21$ TRx locations are positioned along a `T'-shaped measurement route with a uniform spacing of $1$~m between adjacent points, enabling high spatial resolution for geometry-aware analysis. To ensure a fair comparison of material-related channel parameters under consistent geometric conditions, the distance between the TRx and the flat wall is deliberately fixed at $1.2$~m across all three scenarios. Note that this controlled setup isolates the impact of wall material by eliminating distance- and geometry-induced variability, and it does not imply a limitation of the proposed model to fixed-distance configurations.

These scenarios encompass a wide range of material types, including metal, polymer, cement, tile, and glass, along with representative indoor reflector geometries such as flat walls, concave corners, and convex corners. In total, channel measurements are performed at $57$ co-located TRx positions. At each position, $360$ angular measurements are taken, resulting in a total of $57\times360 = 20520$ channel frequency responses (CFRs) across the three indoor environments. These environments exhibit diverse surface materials, structural geometries, and spatial layouts, providing a dataset for THz monostatic channel analysis.

\section{Data Processing and Analysis}\label{sec:estimator}
This section outlines the signal processing approach for the measured data. It starts with the introduction of the signal model for the monostatic sensing case. Using this model, key MPC parameters are extracted through a SAGE-based parameter estimation algorithm. Subsequently, an image-based method is applied to detect and group the MPC clusters.

\subsection{Signal Model}
The directional CFR in the rotation angle of $\theta$ in the $m^{\rm th}$ location can be expressed as
\begin{align}
H_{m}(f,\theta) = \sum_{\ell=1}^{L_m} \alpha_{\ell} \exp(-j2\pi f \tau_{\ell}) \cdot a_{\rm TRx,\phi_{\ell}}(f,\theta),\label{equ:signal}
\end{align}
where $\alpha_{\ell}$, $\tau_{\ell}$, $\phi_{\ell}$, and $f$ denote the complex amplitude (capturing both power attenuation and initial propagation phase), the delay, the azimuth angle of the $\ell^{\text{th}}$ path, and the carrier frequency, respectively. Note that in this work, we assume that the departure azimuth angle and impinge azimuth angle are the same as $\phi_{\ell}$ for the $\ell^{\rm th}$ path. $L_m$ represents the total number of MPCs. $a_{\rm TRx,\phi_{\ell}}(f,\theta)$ is the rotation manifold coefficient, which contains the phase difference caused by DSS and the antenna pattern for $\ell^{\rm th}$ path, and can be written as
\begin{align}
a_{\rm TRx,\phi_{\ell}}(f,\theta) = \left[\exp(j2\pi f r \cos(\theta)/\mathrm{c})\right]^{2} \cdot G_{\rm TRx}(f,\phi_{\ell}-\theta),
\end{align}
where $r$ and $c$ represent the azimuthal distance from the antenna phase center to the rotation center and the speed of light, respectively. $G_{\rm TRx}(f,\phi_{\ell}-\theta)$ is the complex radiation pattern of the mimic monostatic sensor. In this paper, a trajectory-tracking-assisted SAGE algorithm in~\cite{monostatic_lyu} is employed to de-embed the antenna patterns and extract the MPC parameters, i.e., amplitude, delay, and angle, from the measurement data. The extracted parameter set $\hat{\varTheta}^{(m)}$ in the $m^{\rm th}$ location can be expressed as
 \begin{align}
\hat{\varTheta}^{(m)} = \left[ \begin{matrix}
\hat{\alpha}_{1}^{(m)}&	\cdots & \hat{\alpha}_{\ell}^{(m)} &\cdots &  \hat{\alpha}_{Lm}^{(m)}\\
\hat{\tau}_{1}^{(m)}&    \cdots & \hat{\tau}_{\ell}^{(m)} &\cdots &    \hat{\tau}_{Lm}^{(m)}\\
\hat{\phi}_{1}^{(m)}&    \cdots & \hat{\phi}_{\ell}^{(m)} &\cdots &    \hat{\phi}_{Lm}^{(m)}\\
\end{matrix} \right],
 \end{align}
where $\hat{\alpha} _{\ell}^{(m)}$, $\hat{\tau} _{\ell}^{(m)}$, and $\hat{\phi}_{\ell}^{(m)}$ denote the extracted amplitude, delay, and azimuth angle of $\ell^{\rm th}$ MPC, respectively.

\subsection{Image-processing-based Cluster Identification}
After extracting the MPC parameters from the measurement data using SAGE algorithm, an image-processing-based method is used to identify the MPC clusters in the measured power-angle-delay profiles (PADPs) of each TRx location. The measured CFR $H_{m}(f,\theta)$ in the $m^{\rm th}$ location is first transformed to channel impulse response (CIR) $h_m(\tau, \theta)$ through the inverse discrete Fourier transform (IDFT). Then, the PADP $\rm PADP(\tau, \theta)$ can be obtained as
\begin{equation}
\rm PADP(\tau, \theta) = 20\cdot \log_{10} \left( |h(\tau, \theta)| \right).
\end{equation}

The signal region $S(\tau, \theta)$ is first constructed based on the PADP as

\begin{equation}
S(\tau, \theta) =
\begin{cases}
1, & \text{if } \text{PADP}(\tau, \theta) > T \\
0, & \text{otherwise}
\end{cases},
\label{eq:binary_mask}
\end{equation}
where $T$ denotes the signal threshold. In this work, $T$ is set to be $10$~dB above the noise floor~\cite{monostatic_lyu}. 

To enhance the connectivity of detected signal regions and suppress small-scale noise, a morphological closing operation is applied to the binary signal map, which is widely used in the image processing domain~\cite{im_process1,im_process2}. This yields a refined region $S_{\text{closed}}(\tau, \theta)$ defined as:
\begin{equation}
S_{\text{closed}}(\tau, \theta) = \left(S(\tau, \theta) \oplus \mathcal{K}\right) \ominus \mathcal{K},
\label{eq:morph_closing}
\end{equation}
where $\oplus$ and $\ominus$ denote the dilation and erosion operations, respectively, and $\mathcal{K} \subseteq \mathbb{Z}^2$ is a predefined structuring element. Here, $\mathbb{Z}^2$ refers to all $2$D grid points with integer coordinates.

Let $A \subseteq \mathbb{Z}^2$ denote a detected signal region on the delay-angle domain. The dilation operation $\oplus$ expands the boundaries of \( A \), which can be expressed as
\begin{equation}
A \oplus \mathcal{K} = \{ z \in \mathbb{Z}^2 \mid (\mathcal{K}^* + z) \cap A \neq \emptyset \},
\end{equation}
where $z$ is a candidate grid position, and $\mathcal{K}^* = \{ -k \mid k \in \mathcal{K} \}$ denotes the the symmetric of $\mathcal{K}$. The notation $\mathcal{K}^* + z$ represents a spatial shift of $\mathcal{K}^*$ by vector $z$.

In contrast, the erosion operation $\ominus$ shrinks the boundaries of \( A \), removing small, isolated components and smoothing edges, which can be defined as
\begin{equation}
A \ominus \mathcal{K} = \{ z \in \mathbb{Z}^2 \mid (\mathcal{K} + z) \subseteq A \}.
\end{equation}

Subsequently, a connected component labeling (CCL) algorithm is applied to segment contiguous signal regions, where each region is defined as a group of adjacent foreground pixels that share the same label~\cite{gil2002efficient}. To eliminate spurious noise, regions with fewer than $N_{\mathrm{min}}$ pixels are discarded. The $i^{\mathrm{th}}$ delay-angle cluster region $\mathcal{C}_i$ is defined as: 

\begin{figure}
    \centering
    \subfigure []{\includegraphics[width=1.0\columnwidth]{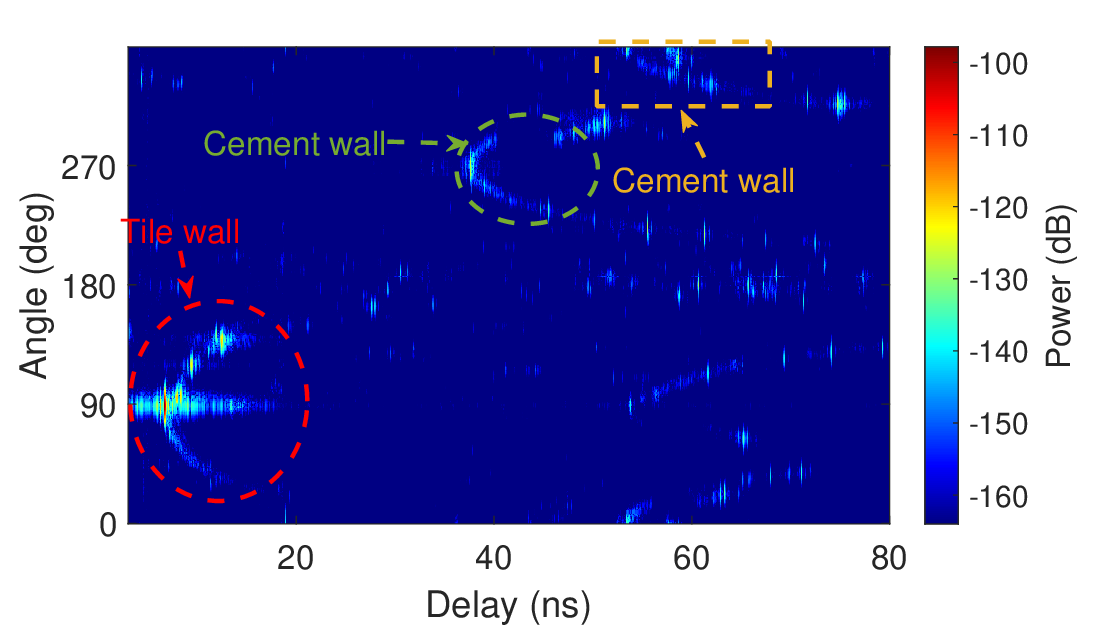}} 
    \subfigure []{\includegraphics[width=1.0\columnwidth]{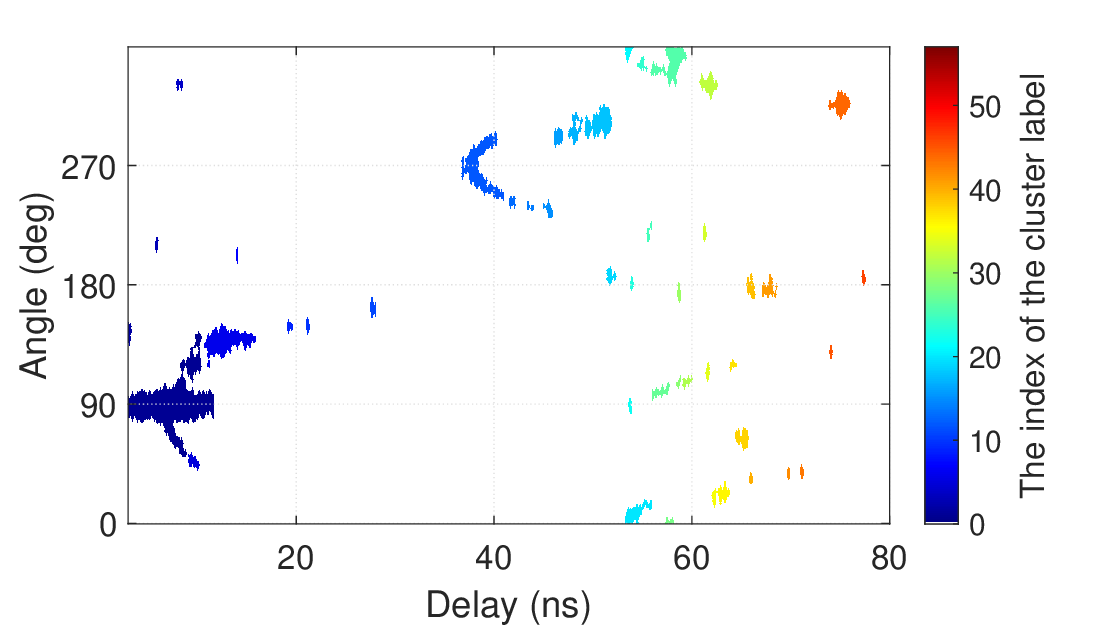}}
    \subfigure []{\includegraphics[width=1.0\columnwidth]{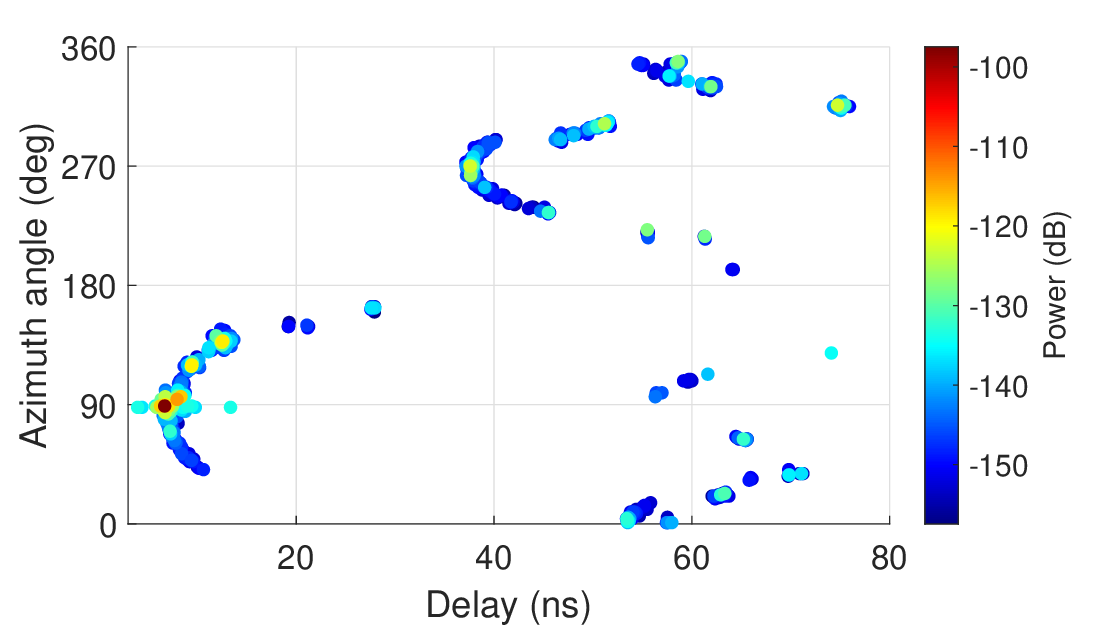}}
\caption {Results of PADP, identified MPC clusters, and the estimated MPCs. (a) Exemplary PADP in TRx $40$. (b) Identified MPC clusters after CCL in TRx $41$. (c) Estimated MPCs by SAGE algorithms.}\label{fig:ccl_example}
\end{figure}

\begin{equation}
\mathcal{C}_i = \left\{ (\tau, \theta) \mid I(\tau, \theta) = i,\; S(\tau, \theta) = 1 \right\},
\label{eq:roi}
\end{equation}
where $I(\tau, \theta)$ denotes the labeled image, which assigns label 
$i$ to the point $(\tau,\theta)$.

Finally, the union of all valid cluster regions can be expressed as
\begin{equation}
\mathcal{C} = \bigcup^{I}_{i} \mathcal{C}_i.
\end{equation}

Fig.~\ref{fig:ccl_example} presents an example of the PADP and the corresponding MPC clusters identified using the CCL algorithm and the estimated MPC after using the SAGE algorithm in \cite{monostatic_lyu}. As shown in Fig.~\ref{fig:ccl_example}~(a), the primary structural features of the scenario, such as the tile and cement walls, are clearly visible. The application of CCL enables effective segmentation of the MPC clusters, as illustrated in Fig.~\ref{fig:ccl_example}~(b). Besides, the MPCs are accurately estimated from the measured CFR data, as depicted in Fig.~\ref{fig:ccl_example}~(c).

\section{Cluster-Based Channel Characterization}\label{sec:modeling}
This section analyzes the channel characteristics of monostatic sensing at the $300$~GHz band across the three indoor scenarios. In this context, a cluster denotes a group of MPCs exhibiting relative proximity in delay and angular domains, generally resulting from specular and diffuse reflections off one or more closely spaced reflectors~\cite{cluster_def1,cluster_def2,cluster_def3}. These MPCs are perceived as a coherent group in the delay-angle domain and collectively characterize the channel response associated with a local scattering region in the environment.

\subsection{Identification of the Specular and Diffuse Components}\label{sec:idenfify}
In monostatic sensing channels, specular and diffuse reflections constitute the primary propagation mechanisms. Reflection loss serves as a distinguishing parameter between specular and diffuse components within MPC clusters, which are predominantly composed of specular reflections. In our analysis, specular components are identified by their relatively lower reflection losses and alignment with strong, geometrically consistent propagation paths, whereas diffuse components exhibit higher losses and broader angular spreads. A threshold is selected based on the statistical distribution of reflection losses observed across the dataset. The reflection loss $Loss_{\ell}$ for $\ell^{\rm th}$ de-embedded MPC can be expressed as
\begin{align}
Loss_{\ell} [\rm dB] =  \rm FSPL_{\ell} [\rm dB] - 20\cdot \log_{10}|\hat{\alpha}_{\ell}|,
\end{align}
where $\rm FSPL_{\ell}$ (in decibels) is the free space path loss through the propagation channel. The free space path loss $\rm FSPL_{\ell}$ can be calculated as
\begin{align}
{\rm FSPL}_{\ell} = 20\cdot \log_{10}\left( \frac{4\pi f_{c} d_{\ell}}{\mathrm{c}}\right),
\end{align}
with $f_{c}$ denoting the centering frequency, and $d_{\ell} = \hat{\tau}^{(m)}_{\ell}\cdot c$ representing the round trip propagation distance of $\ell^{\rm th}$ path. 

\begin{figure}
    \centering
    \subfigure []{\includegraphics[width=1.0\columnwidth]{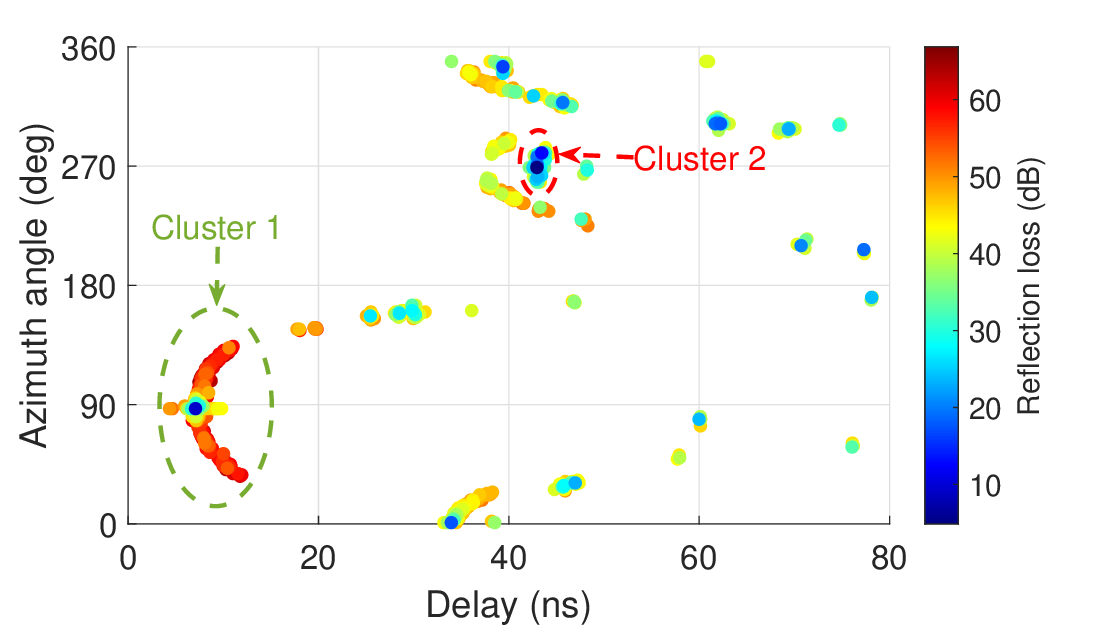}} 
    \subfigure []{\includegraphics[width=0.8\columnwidth]{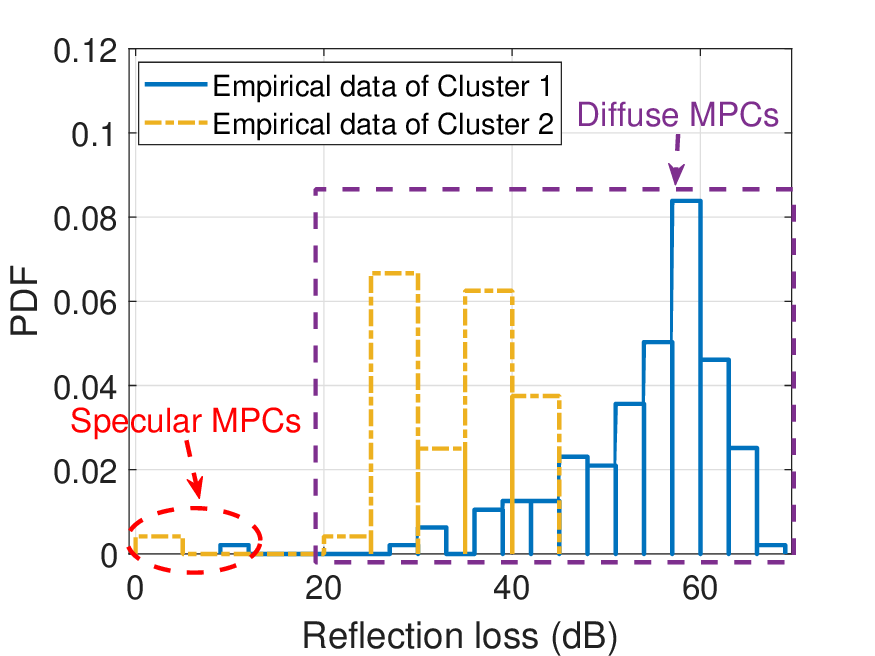}}
\caption {Exemplary results of reflection loss analysis. (a) The comparison of the reflection loss and MPC distribution at TRx~$41$. (b) Probability density function (PDF) of the reflection loss for MPC clusters at TRx~$41$.}\label{fig:refl_loss}
\end{figure}

The exemplary reflection loss results are shown in Fig.~\ref{fig:refl_loss}. As illustrated in Fig.~\ref{fig:refl_loss}~(a), each identified cluster typically contains a dominant specular MPC characterized by low reflection loss, along with multiple diffuse MPCs exhibiting higher reflection losses. The probability density functions (PDFs) of reflection loss for Clusters~$1$ and $2$ are shown in Fig.\ref{fig:refl_loss}(b). In Cluster~$1$, the specular MPC exhibits the lowest reflection loss of $2.5$~dB, while the reflection losses of diffuse MPCs span a wide range from $27$ to $69$~dB. In Cluster~$2$, the reflection loss of the specular MPC is $10.49$~dB, and that of the diffuse MPCs ranges from $20$ to $45$~dB. These results provide physical insight into the underlying propagation mechanisms, i.e., specular MPCs arise from strong, geometrically aligned reflections with minimal scattering loss, while diffuse MPCs are caused by interactions with rough or irregular surfaces that scatter energy over wider angular ranges, resulting in higher reflection losses. The clear separation in reflection loss between specular and diffuse components indicates that reflection loss is a reliable and interpretable metric for classifying multipath components in THz monostatic sensing channels.

\subsection{Number of MPC Clusters}
\begin{figure}
\centering
\includegraphics[width=0.8\columnwidth]{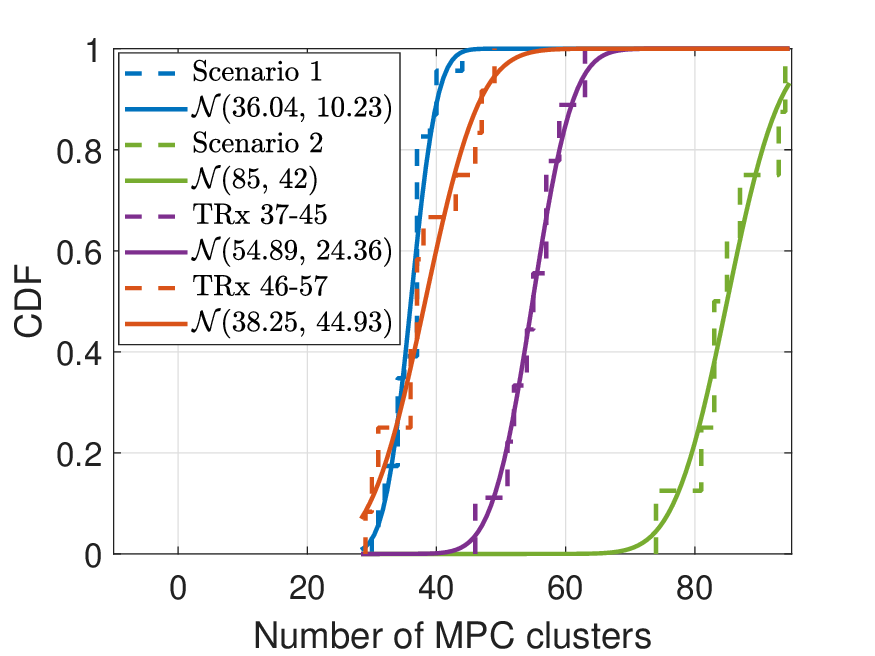}
\caption {The CDF of the number of MPC clusters and the fitted models.}\label{fig:num_clusters}
\end{figure}

The number of MPC clusters serves as an important metric for analyzing the complexity of a propagation environment. In this work, the number of MPC clusters is examined across three indoor scenarios. Specifically, in Scenario~$3$, TRxs $37$-$45$ and TRxs $46$-$57$ are treated as two distinct cases. The cumulative distribution functions (CDFs) of the number of MPC clusters are presented in Fig.~\ref{fig:num_clusters}. It can be observed that Scenario~$1$ and TRxs $37$-$45$ exhibit a similar number of MPC clusters. TRxs $37$-$45$ is shown to have a higher number of clusters compared to those in Scenario~$1$ and TRxs $37$-$45$. In contrast, Scenario~$2$ has the highest number of MPC clusters among all cases, which can be attributed to the presence of more metallic structures that act as strong reflectors. The CDFs of the number of clusters in the four cases align well with normal distributions, characterized by the parameters $\mathcal{N}(36.01, 10.23)$, $\mathcal{N}(85, 42)$, $\mathcal{N}(54.89, 24.36)$, and $\mathcal{N}(38.25, 44.93)$, respectively, as illustrated in Fig.~\ref{fig:num_clusters}.

\subsection{Intra-Cluster Delay and Angular Characteristics}
The channel characteristics in the delay and angle domains are essential for characterizing the shape of MPC clusters, which in turn reflect the size and geometry of reflectors present in the environment. In this work, we characterize cluster parameters using four key intra-cluster metrics: delay depth, angular width, delay spread, and angular spread.

\begin{figure}
    \centering
    \subfigure []{\includegraphics[width=0.8\columnwidth]{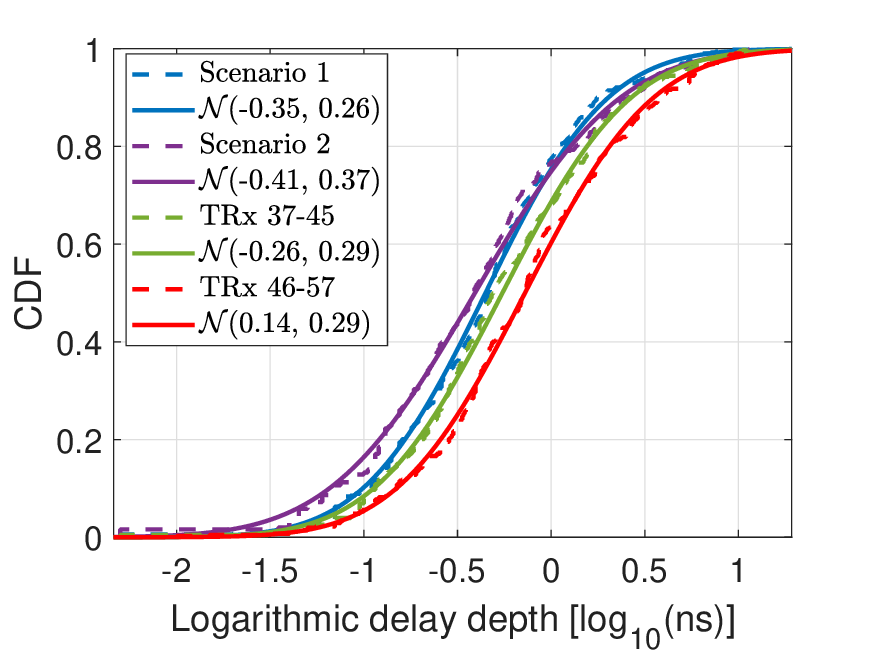}} 
    \subfigure []{\includegraphics[width=0.8\columnwidth]{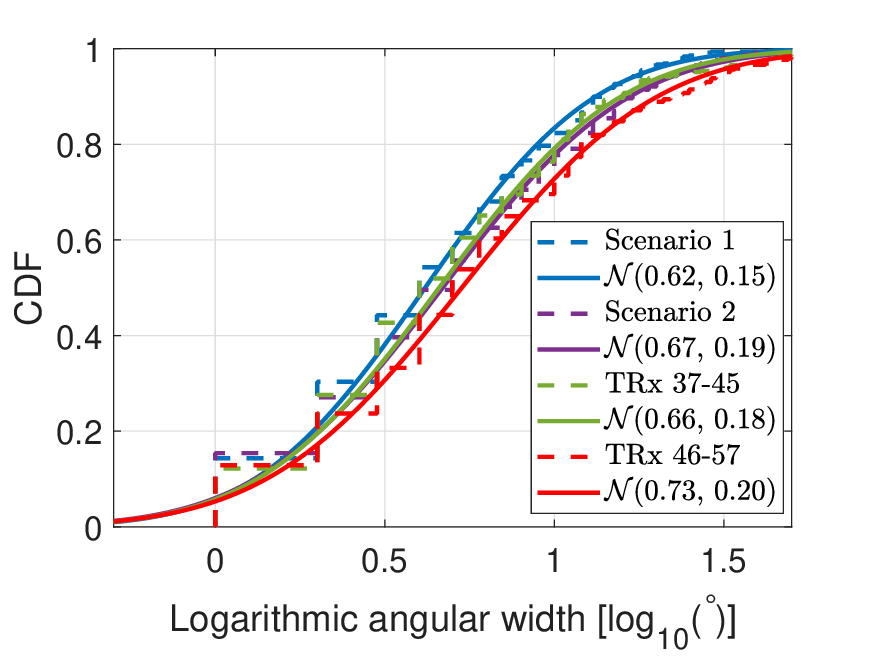}}
\caption {The CDF of the logarithmic delay depth, logarithmic angular width, and the fitted models. (a) Delay depth. (b) Angular width.}\label{fig:depth_and_width}
\end{figure}

In this work, delay depth is defined as the total delay range between the earliest and latest arrived MPCs within a cluster, describing the full temporal extent of the cluster. Angular width refers to the difference between the minimum and maximum angles of arrival of the MPCs in the cluster, representing its overall angular coverage. Fig.~\ref{fig:depth_and_width} illustrates the CDF of the delay depth, angular width, and the corresponding fitted models. For the delay depth, the four cases exhibit the ranges of $[0.005,17.73]$~ns, $[0.01,12.53]$~ns, $[0.005,11.03]$~ns, and $[0.01,10.74]$~ns with the corresponding average values of $0.89$, $1.00$, $1.16$, and $1.47$~ns, respectively. As shown in Fig.~\ref{fig:depth_and_width}~(a), the CDFs of the delay depths for the four cases, plotted on a logarithmic scale, closely follow normal distributions. In terms of angular width, the four cases demonstrate a range of values spanning $[1^{\circ},136^{\circ}]$, $[1^{\circ},120^{\circ}]$, $[1^{\circ},104^{\circ}]$, and $[1^{\circ},117^{\circ}]$, respectively. The corresponding average angular widths are $8.83^{\circ}$, $9.71^{\circ}$, $8.28^{\circ}$, and $10.81^{\circ}$, respectively. In Fig.~\ref{fig:depth_and_width}~(b), it can be found that the logarithmic angular widths in these four cases are fitted well with the normal distributions.

\begin{figure}
    \centering
    \subfigure []{\includegraphics[width=0.8\columnwidth]{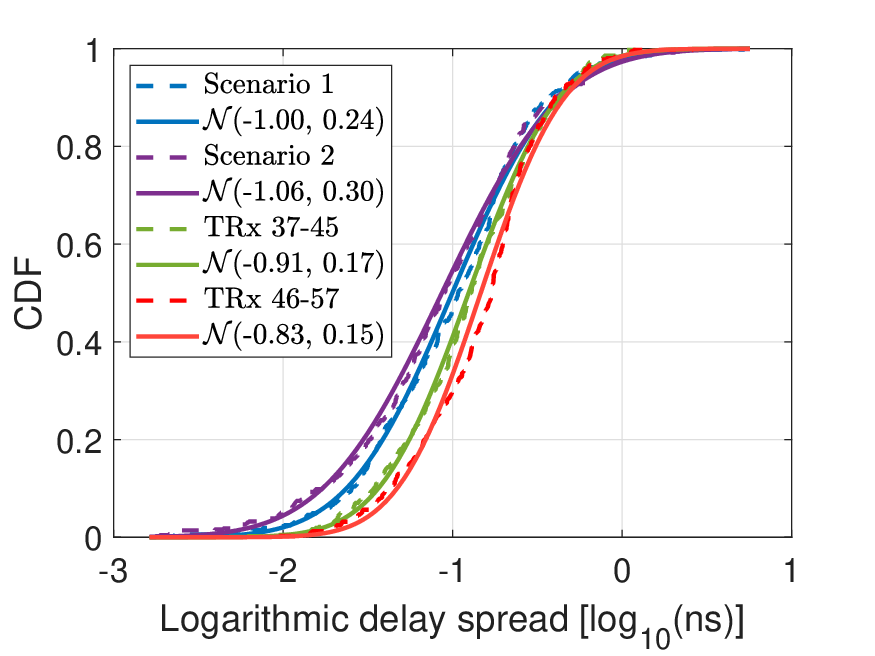}} 
    \subfigure []{\includegraphics[width=0.8\columnwidth]{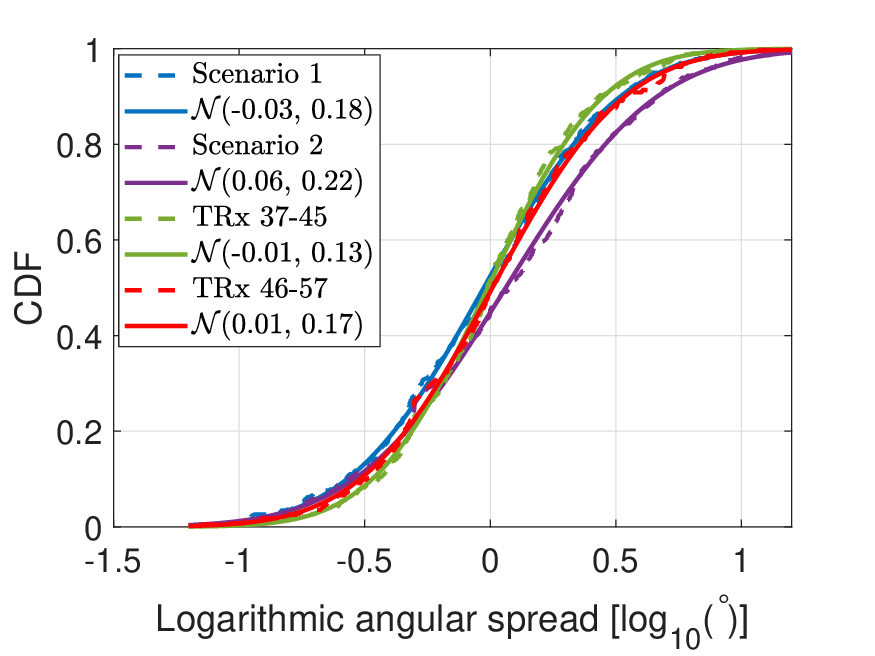}}
\caption {The CDF of the logarithmic delay spreads, logarithmic angular spreads, and the fitted models. (a) Delay spreads. (b) Angular spreads.}\label{fig:spreads}
\end{figure}

The intra-cluster delay and angular spreads are key parameters used to characterize the dispersion of MPCs within a cluster in the delay and angle domains, respectively. The delay and angular spreads characterize the statistical dispersion of MPCs within a cluster. In contrast, delay depth and angular width refer to the absolute range of the cluster in the delay and angle domains. Both metrics provide complementary insights, i.e., depth and width reflect the geometric extent of MPCs in the channel, while spreads capture the distribution of energy, which is critical for accurate channel modeling and ISAC system design. Detailed definitions and formulations can be found in~\cite{del_spread,aoa_spread,3GPP_38.901} for the detailed explanation and equations of the delay and angular spreads. Fig.~\ref{fig:spreads} presents the CDFs of delay and angular spreads, along with their respective fitted distribution models. Regarding the intra-cluster delay spread, Scenario~$2$ exhibits the smallest spread, with an average value of $0.17$~ns. In comparison, Scenario~$1$ and TRxs~$37$-$45$ both show slightly larger average spreads of $0.18$~ns. TRxs~$46$-$57$ display the largest intra-cluster delay spread among the four cases, with a mean value of $0.21$~ns. The corresponding ranges of delay spreads are $[0,2.94]$, $[0,5.24]$, $[0,1.20]$, and $[0,2.60]$~ns, respectively. Similar to the analysis of delay depth and angular width, normal distributions are fitted to the logarithmic intra-cluster delay spreads, showing a strong agreement between the empirical data and the fitted models, as depicted in Fig.~\ref{fig:spreads}~(a). In terms of the intra-cluster angular spreads, the angular spreads of these four cases show a very close values, with the range of $[0^{\circ},10.78^{\circ}]$, $[0^{\circ},25.65^{\circ}]$, $[0^{\circ},13.64^{\circ}]$, and $[0^{\circ},22.56^{\circ}]$, respectively. The corresponding mean angular spreads are $1.40^{\circ}$, $1.71^{\circ}$, $1.27^{\circ}$, and $1.48^{\circ}$, respectively. In Fig.~\ref{fig:spreads}(b), normal distribution fitting is also applied to the CDF of the logarithmic intra-cluster angular spreads, showing a good match between the fitted model and the empirical data.

\begin{table*}[t]
\centering
\caption{Summary of the Proposed Sensing Channel modeling framework}
\label{tab:feature_summary}
\begin{tabular}{p{0.8cm} p{3cm} p{5.5cm} p{6cm}}
\hline
\textbf{Level} & \textbf{Physical Feature} & \textbf{Key Propagation Observables} & \textbf{Main Output} \\
\hline
1 & Reflector quantity \& location & Number of MPC clusters \& cluster delay and angle centroids from high-resolution parameter estimation & Spatial distribution and count of dominant reflectors \\
\hline
2 & Surface roughness & Number of diffuse MPCs within each cluster \& Lambertian model & Relative roughness classification and scattering strength \\
\hline
3 & Reflector geometry & Spatial distribution of MPCs \& intra-cluster delay-angle dispersions;  & Geometric categorization of reflectors \\
\hline
4 & Material type & Specular and diffuse reflection loss distributions & Material identification and EM property estimation \\
\hline
\end{tabular}
\end{table*}

\section{Environment-Aware Sensing Channel Modeling Framework}\label{sec:material}
\begin{figure}
\centering
\includegraphics[width=1\columnwidth]{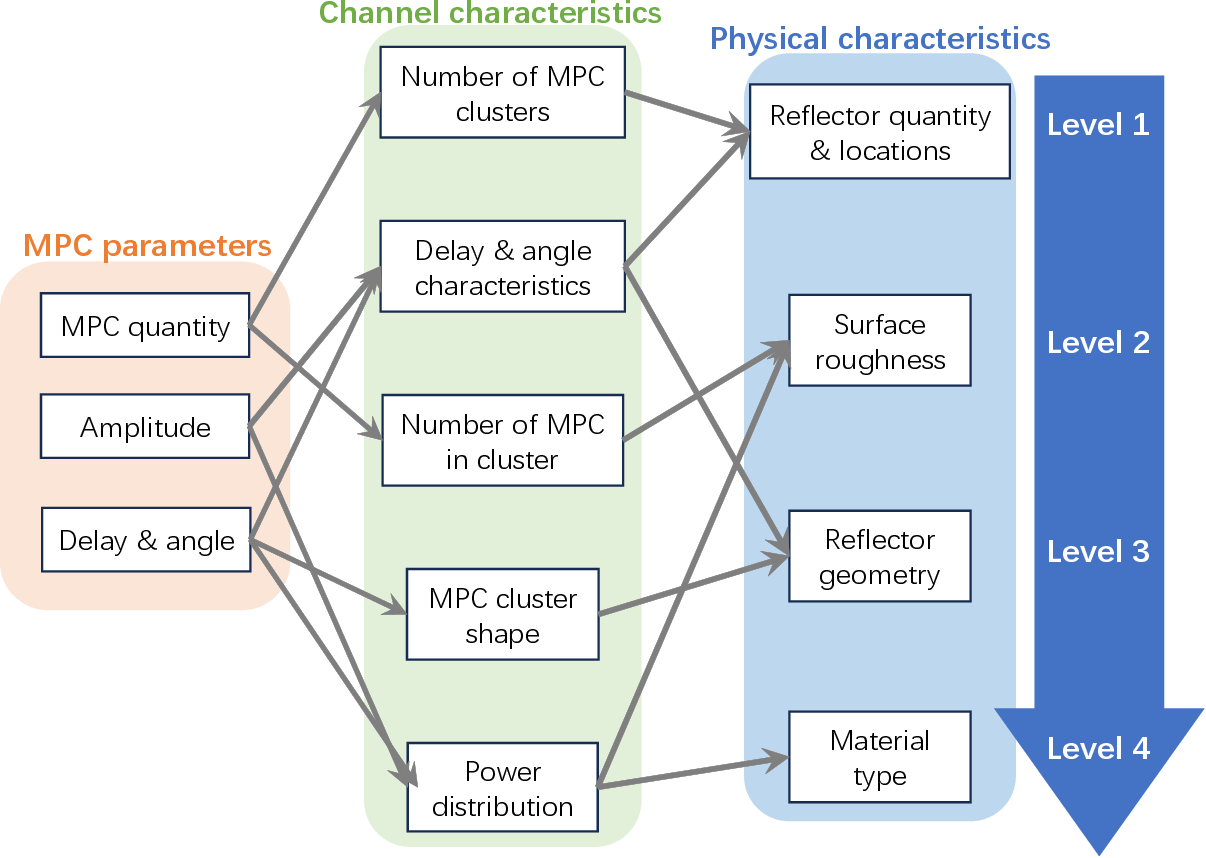}
\caption {The framework of the environment-aware channel modeling for monostatic sensing.}\label{fig:framework}
\end{figure}

This section introduces an environment-aware channel modeling framework for monostatic sensing, which establishes a mapping between the physical characteristics of the environment and the channel features. Fig.~\ref{fig:framework} illustrates the proposed environment-aware channel modeling framework. Besides, a summary of the proposed framework, linking measurable channel observables to inferred physical attributes, is provided in Table~\ref{tab:feature_summary}. Specifically, the quantity and spatial distribution of reflectors influence the number of MPC clusters and their delay-angle information, as more complex environments tend to generate a greater number of distinguishable reflection groups. Surface roughness impacts both the number and angular span of diffuse MPCs within each cluster, i.e., rougher surfaces scatter energy more broadly. Besides, the surface roughness can be characterized using power distribution in the angle domain, i.e., the Lambertian model. Reflector geometry further shapes the spatial arrangement of MPCs, influencing cluster boundaries and intra-cluster dispersion patterns, depending on the physical layout (e.g., a flat wall or a corner). Material type are manifested in the reflection loss distribution between specular and diffuse components, which is governed by the electromagnetic (EM) properties of different materials. Note that in this work, the TRx-to-reflector distance is fixed at $1.2$~m across all scenarios to maintain consistent geometric conditions when isolating material-related effects. In practical sensing deployments, the TRx-reflector distance may vary or be unknown; however, its influence on the derived parameters can be calibrated or compensated for during processing (e.g., in reflection loss estimation). While the Lambertian scattering model is generally stable under small distance variations, larger changes in range may alter the effective scattering geometry, which warrants further investigation in future work. The fixed-distance configuration adopted here, therefore, serves as a controlled reference for subsequent studies considering variable-distance scenarios.

These four physical features can be ranked by the level of inference difficulty in THz monostatic sensing. The easiest to extract is the reflector quantity and location (Level~$1$), which can be reliably identified using high-resolution parameter estimation algorithms that extract delay and angle information of MPC clusters~\cite{yuanbo_sage,sage_algo}. Surface roughness (Level~$2$) follows, as the short wavelengths at THz frequencies make the system highly sensitive to fine surface textures. This enables the detection of scattering behavior associated with rough surfaces, as demonstrated in several recent studies~\cite{roughness_analysis1,roughness_analysis2,roughness_analysis3}. Reflector geometry (Level~$3$) poses a greater challenge, as it requires spatial interpretation of the shape and angular distribution of clustered diffuse MPCs. However, the pronounced surface scattering at THz frequencies, resulting from sub-millimeter wavelengths, makes it possible to infer reflector shapes such as corners and curved surfaces~\cite{monostatic_lyu}. The most difficult task is the identification of material type (Level~$4$), which involves distinguishing between different materials based on their specular and diffuse reflection behavior. This requires accurate statistical modeling, which forms a key contribution of this work.

By explicitly connecting these hierarchical physical attributes to observable propagation features, the proposed environment-aware modeling framework enables more accurate and interpretable THz channel representations tailored for sensing-aware applications. The following subsections provide a detailed investigation of key material-related features, namely, surface roughness, reflector geometry, and material composition.

\begin{figure}
    \centering
    \subfigure []{\includegraphics[width=0.8\columnwidth]{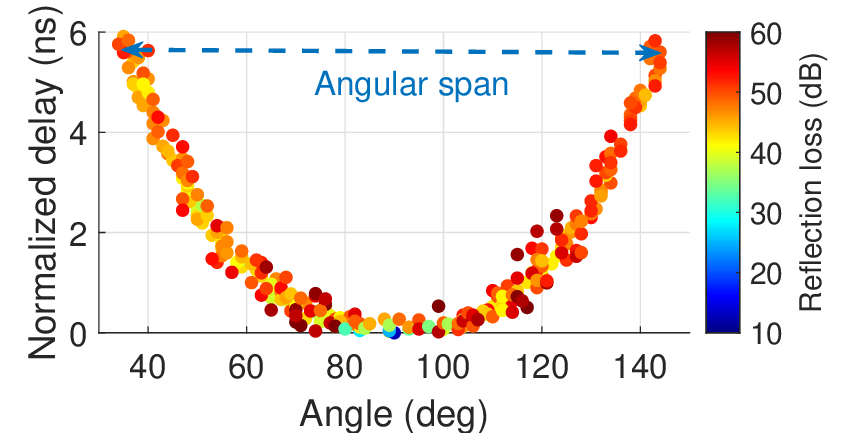}} 
    \subfigure []{\includegraphics[width=0.8\columnwidth]{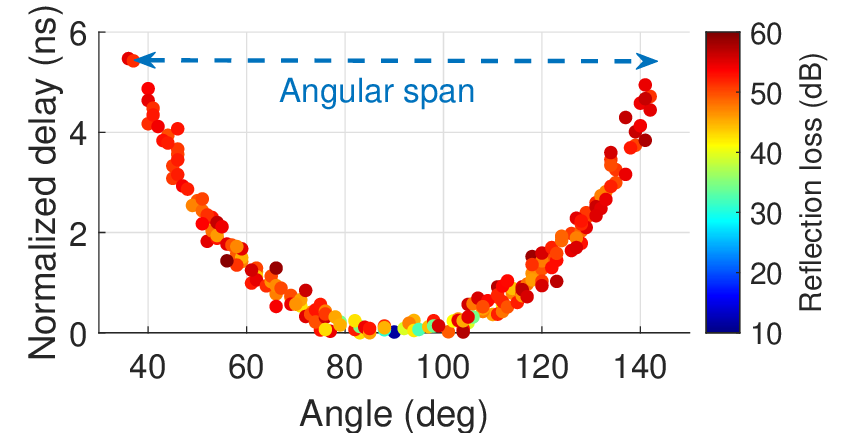}}
    \subfigure []{\includegraphics[width=0.8\columnwidth]{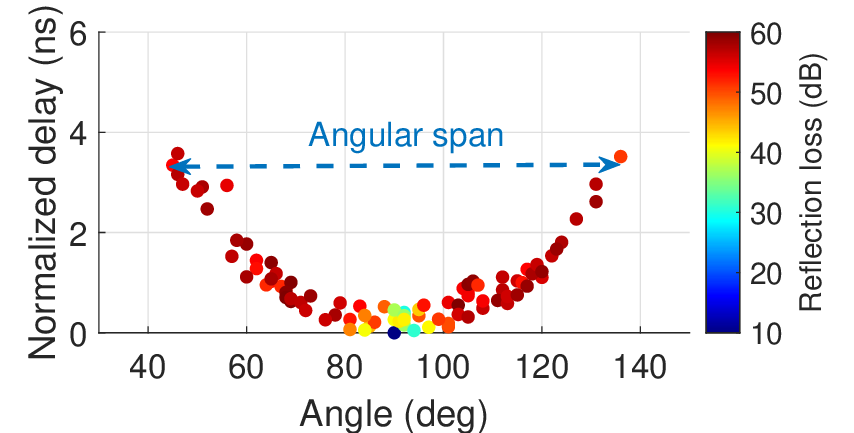}}
\caption {The reflected MPCs from flat walls composed of three materials with varying roughness. (a) Polymer. (b) Cement. (c) Tile.}\label{fig:roughness_padp}
\end{figure}

\begin{figure}
    \centering
    \subfigure []{\includegraphics[width=0.8\columnwidth]{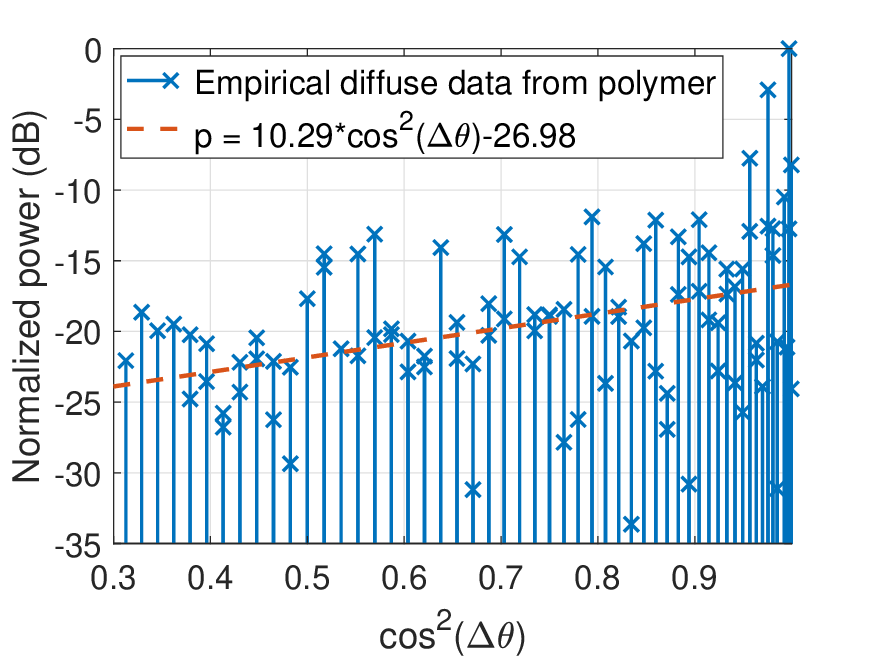}} 
    \subfigure []{\includegraphics[width=0.8\columnwidth]{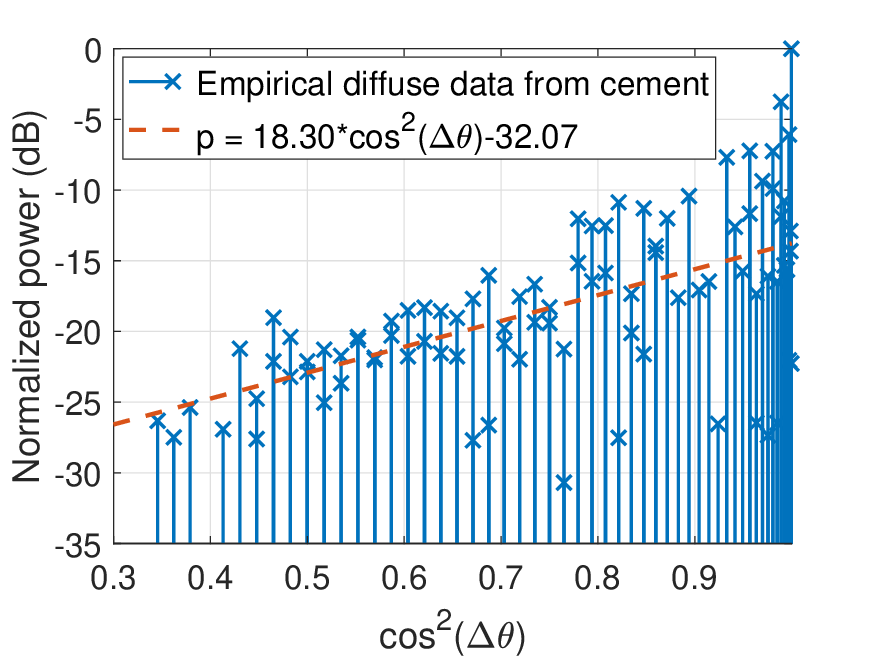}}
    \subfigure []{\includegraphics[width=0.8\columnwidth]{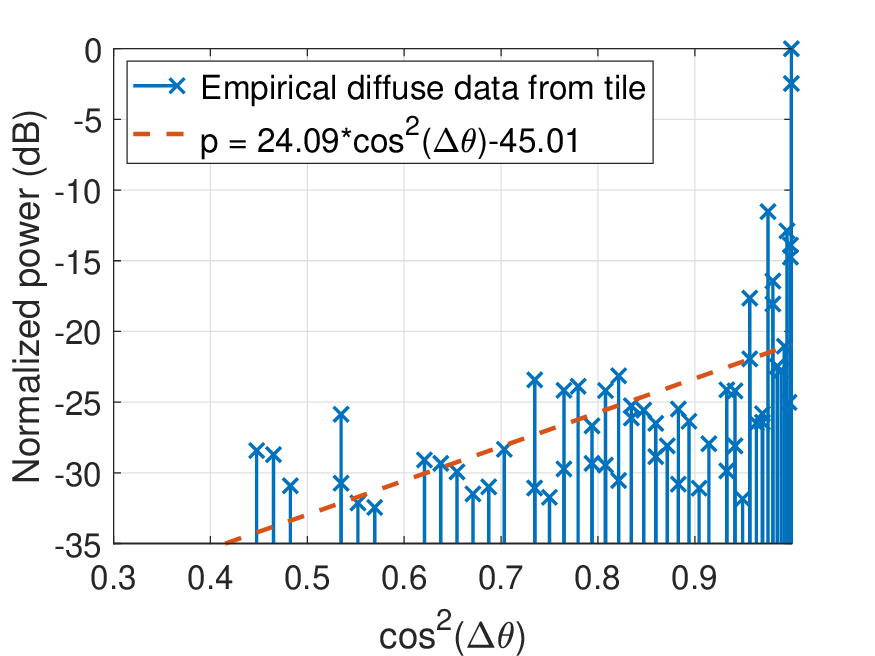}}
\caption {The relationship between the diffuse power with $\cos^{2}(\Delta  \theta)$ and the fitted Lambertian model. (a) Polymer. (b) Cement. (c) Tile.}\label{fig:lambertian_padp}
\end{figure}

\subsection{Surface Roughness}
Surface roughness is a crucial physical property of reflectors, particularly at THz frequencies, where the wavelength is on the order of millimeters. At such short wavelengths, even small surface irregularities can significantly affect wave scattering, making THz signals susceptible to surface texture. Roughness can be quantified by the root mean square (RMS) height $\sigma$ of the surface profile~\cite{roughness_height}, defined as  
\begin{equation}
\sigma = \sqrt{\frac{1}{N}\sum_{i=1}^{N} \left(h_i - \bar{h}\right)^2},
\end{equation}
where $h_i$ is the measured height at the $i$-th point, $\bar{h}$ is the mean height, and $N$ is the total number of sampled points.  

In practical channel measurements, obtaining $\sigma$ is often infeasible in non-invasive scenarios. Thus, in this work, we employ two empirically driven indicators to infer roughness effects indirectly: (i) the number of detected MPCs within a cluster and (ii) the angular spread of reflections modeled using Lambertian-like behavior. To illustrate the first indicator, Fig.~\ref{fig:roughness_padp} presents delay-angle profiles for reflections from flat walls made of polymer, cement, and tile materials with progressively smoother surfaces based on visual inspection and construction standards. The observed angular spans are $110^{\circ}$ (polymer), $106^{\circ}$ (cement), and $90^{\circ}$ (tile), and the corresponding numbers of detected MPCs are $267$, $195$, and $99$, respectively. These results reveal a clear trend: surfaces with greater roughness tend to produce broader angular dispersion and a larger number of diffuse multipath reflections. 

The Lambertian model is also a metric to analyze the surface roughness. According to the Lambertian model~\cite{diffuse_model1}, the diffuse power $p_{\rm Lam}(\Delta \theta)$ is proportional to $\cos^{2}(\Delta \theta)$
\begin{align}
p_{\rm Lam}(\Delta \theta) = n_{\rm Lam} \cdot \cos^{2}(\Delta \theta) + b_{\rm Lam},
\end{align}
where $\Delta \theta$, $n_{\rm Lam}$, and $b_{\rm Lam}$ are the angle difference between the diffuse angle $\theta_{\mathrm{diff}}$ and the specular reflection angle $\theta_{\mathrm{spec}}$, i.e., $\Delta \theta = \theta_{\mathrm{spec}} - \theta_{\mathrm{diff}}$, the slope of the diffuse power compared to $\cos^{2}(\Delta  \theta)$, and the intercept. Note that the diffuse power $p_{\rm Lam}(\Delta \theta)$ is obtained by summing the powers of all diffuse MPCs arriving at the same angle $\theta_{\mathrm{diff}}$.

Fig.~\ref{fig:lambertian_padp} illustrates the comparison between the normalized diffuse power and the corresponding fitted Lambertian scattering models for three surfaces with varying roughness. It can be observed that the diffuse power generally increases with $\cos^{2}(\Delta \theta)$, consistent with the Lambertian scattering principle. The fitted Lambertian models for the polymer, cement, and tile surfaces are $p_{\rm Lam}(\Delta \theta) = 10.29 \cdot \cos^{2}(\Delta \theta) - 26.98$~dB, $p_{\rm Lam}(\Delta \theta) = 18.30 \cdot \cos^{2}(\Delta \theta) - 32.07$~dB, and $p_{\rm Lam}(\Delta \theta) = 24.09 \cdot \cos^{2}(\Delta \theta) - 45.01$~dB, respectively. These results reveal a clear trend: as the surface roughness increases, the slope parameter $n_{\rm Lam}$, representing the degree of angular concentration in the Lambertian fit, decreases. This implies that rougher surfaces yield more widely dispersed diffuse scattering, while smoother surfaces concentrate the diffuse energy more narrowly around the specular direction.

Note that the surface roughness classification in this study is primarily based on empirical observations rather than precise surface measurements. Nevertheless, it provides a practical and intuitive interpretation that aligns with expected surface scattering behavior at high frequencies. We believe such insights are valuable for guiding future model development and measurement-based material classification efforts in THz sensing systems.

\subsection{Reflector Geometry}
In this subsection, two key parameters, i.e., cluster shape and material-specific channel dispersion, are analyzed, as these parameters reflect the geometry and structural characteristics of the reflectors.

\begin{figure}
\centering
\includegraphics[width=1.0\columnwidth]{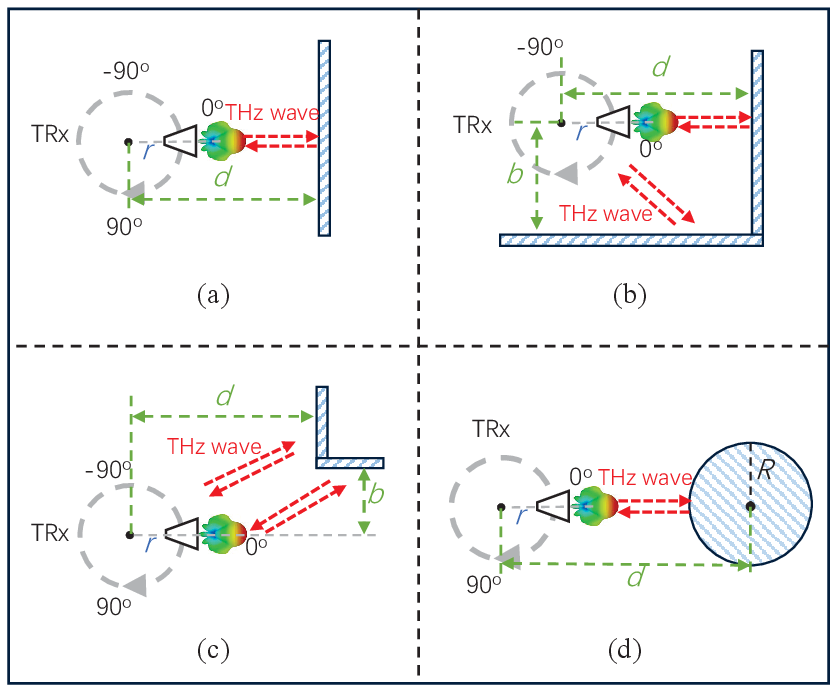}
\caption {The layout of the indoor representative physical structures. (a) Flat wall. (b) Concave corner. (c) Convex corner (d) Cylinder pillar.}\label{fig:structure}
\end{figure}

\begin{figure}
\centering
\includegraphics[width=1.0\columnwidth]{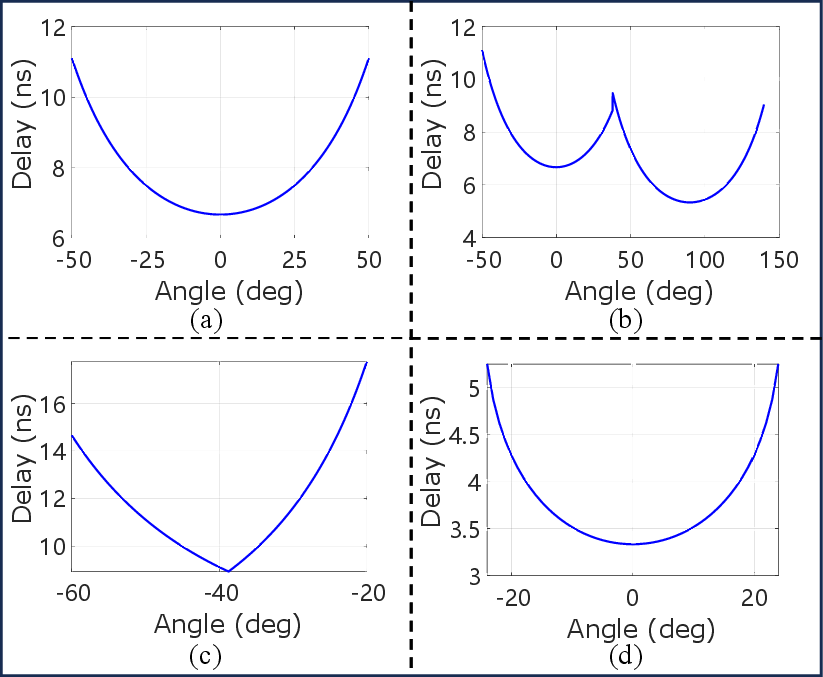}
\caption {The simulated structural model of different physical structures: (a) Flat wall. (b) Concave corner. (c) Convex corner (d) Cylinder pillar.}\label{fig:channel_domain}
\end{figure}

\begin{figure}
    \centering
    \subfigure []{\includegraphics[width=0.85\columnwidth]{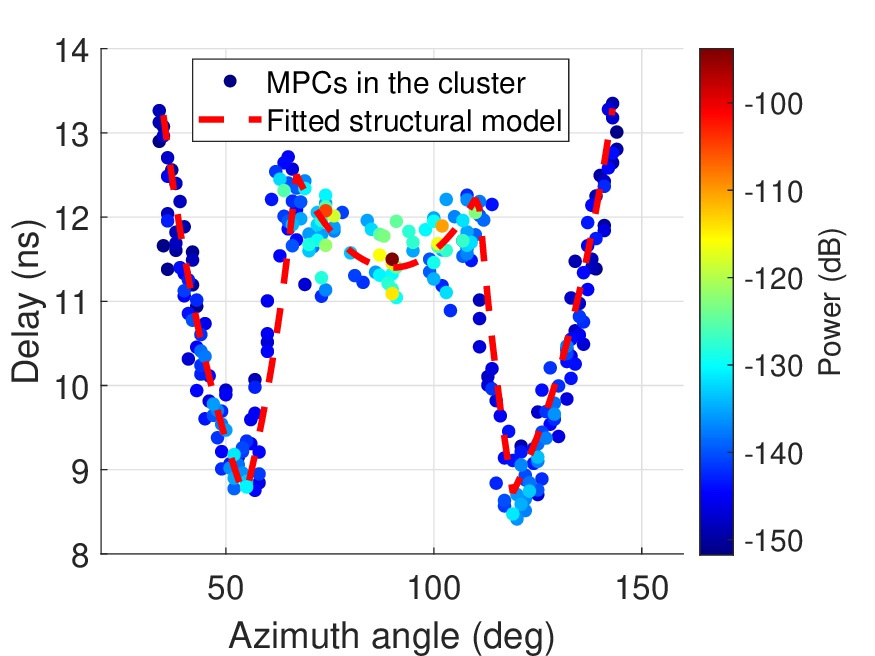}} 
    \subfigure []{\includegraphics[width=0.85\columnwidth]{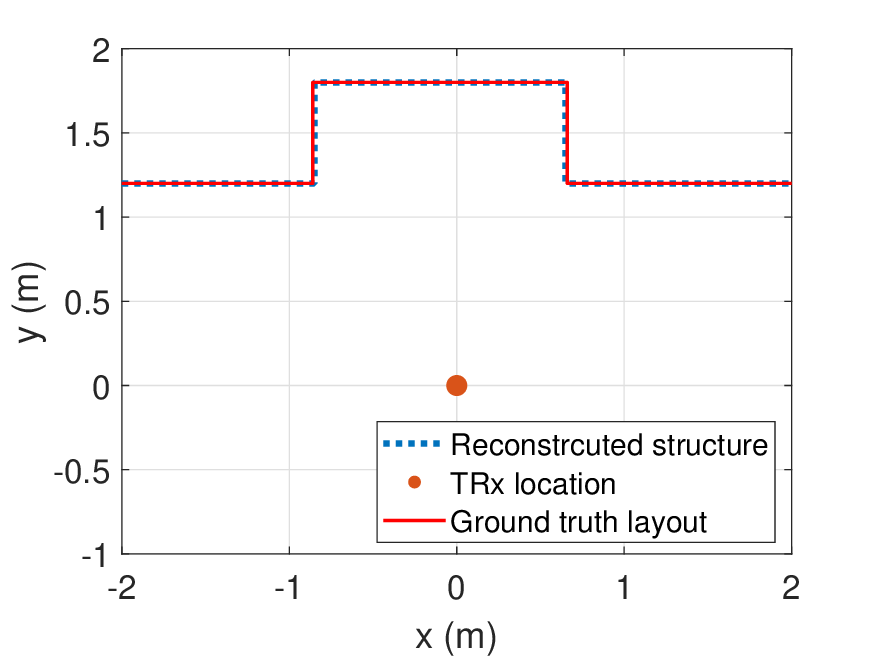}}
\caption {Analysis of structural modeling using empirical data. (a) Example of matching a diffuse MPC cluster with the fitted structural model; (b) Reconstructed environment based on the identified structures.}\label{fig:structural_model}
\end{figure}

\subsubsection{Cluster Shape Analysis}
The diffuse MPCs within a cluster can be effectively utilized to analyze the structural characteristics of reflectors. By examining the spatial distribution and geometric shape of these components, it becomes possible to infer the underlying physical structure of the reflecting objects. In this subsection, we investigate several representative physical structures, namely, a flat wall, a concave corner, a convex corner, and a cylindrical pillar, and analyze their corresponding manifestations in the channel as illustrative examples. The layout of these structures is depicted in Fig.~\ref{fig:structure}. In the simulation setup, same as the measurements, the TRx moves along a circular trajectory with a radius of $r = 0.2$~m relative to the rotation center, and the distance to the reflectors is fixed at $d=1.2$~m. For both the concave and convex corner scenarios, the distance $b$ is set to $1.0$~m. In the cylindrical pillar case, the pillar radius is configured as $R=0.5$~m. All other simulation parameters are aligned with the real measurement settings to ensure consistency. Fig.~\ref{fig:channel_domain} presents the simulation results, showing how these different physical structures are manifested in the delay–angle domain. It can be clearly observed that each structure exhibits distinct channel characteristics, allowing for effective differentiation of reflector shapes based on their channel signatures.

By matching a structural model with the diffuse MPCs within a cluster, the structural characteristics of the reflector can be effectively identified. As an illustrative example, Fig.~\ref{fig:structural_model} presents the matching results for TRx~$50$, showing how the diffuse MPCs of a single cluster align with a fitted structural model. As depicted, two convex corners and one flat wall are identified, which correspond accurately to the actual layout in the measurement environment, as shown in the reconstructed structure in Fig.~\ref{fig:structural_model}~(b). A comparison between the fitted structural model and the ground-truth geometry of Scenario~$3$ confirms a strong agreement, demonstrating the effectiveness of the model in capturing the structural features of the reflectors.

\subsubsection{Material-Specific Channel Dispersion}
The intra-cluster delay and angular spreads can serve as indirect indicators of reflector geometry by capturing the spatial dispersion of multipath components resulting from interactions with physical surfaces. To analyze this, clusters associated with specific reflector materials are identified based on scenario geometry and specular reflection loss. Only clusters involving reflections from a single material type are considered. 

\begin{figure}
    \centering
    \subfigure []{\includegraphics[width=0.8\columnwidth]{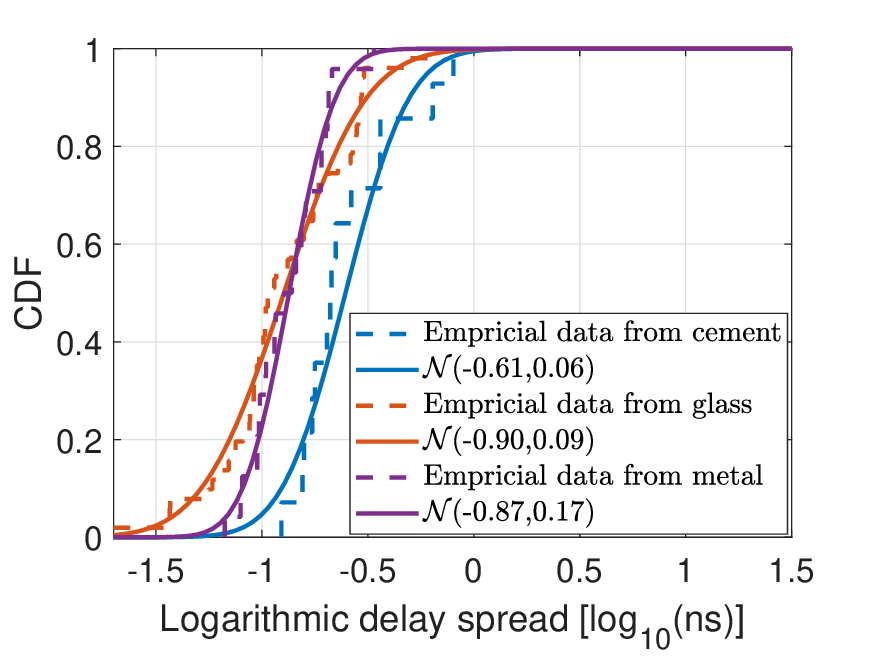}} 
    \subfigure []{\includegraphics[width=0.8\columnwidth]{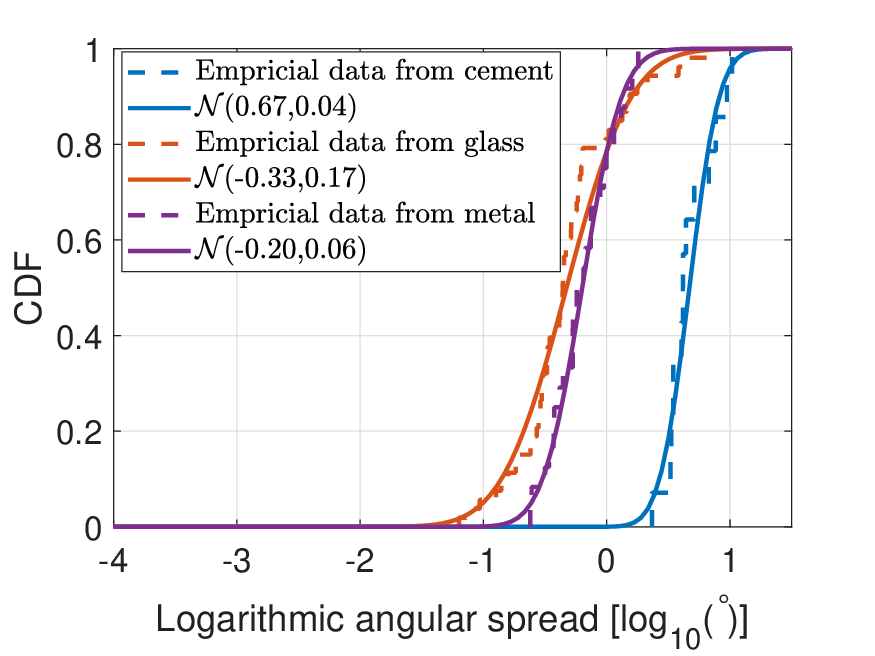}}
\caption {The CDF of the material-specific intra-cluster delay and angular spreads, and the corresponding fitted models. (a) Delay spread. (b) Angular spread.}\label{fig:material_specific}
\end{figure}

Using Scenario~$1$ as an example, Fig.~\ref{fig:material_specific} presents the material-specific intra-cluster spreads. The delay spread ranges for cement, glass, and metal are observed as $[0.12,0.80]$~ns, $[0.004,0.66]$~ns, $[0.07,0.40]$~ns, respectively. In terms of logarithmic angular spread, metal exhibits the narrowest range  $[0^{\circ},1.80^{\circ}]$,, consistent with its small, localized structures (e.g., metallic window frames). Glass surfaces show a moderately wider spread $[0^{\circ}, 7.79^{\circ}]$, while cement surfaces yield the broadest spread $[2.33^{\circ}, 10.41 ^{\circ}]$, reflecting their larger physical extent and rougher texture, which induce wider angular dispersion. These trends suggest that larger reflectors produce more diffuse scattering, contributing to increased intra-cluster spreads. Moreover, the CDFs of both delay and angular spreads exhibit good agreement with fitted normal distributions, as shown in Fig.~\ref{fig:material_specific}.

\begin{figure}
    \centering
    \subfigure []{\includegraphics[width=0.8\columnwidth]{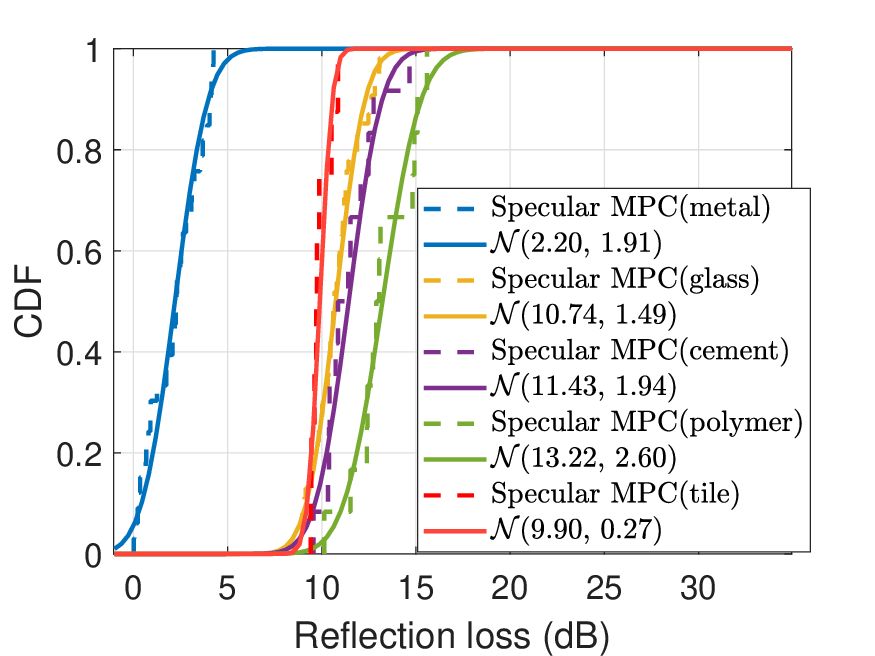}} 
    \subfigure []{\includegraphics[width=0.8\columnwidth]{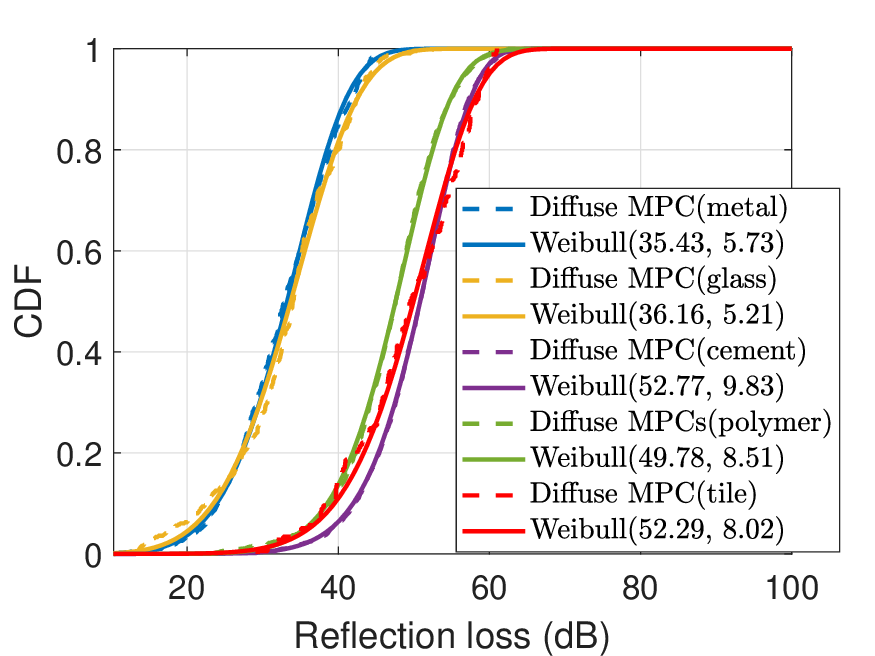}}
\caption {The CDF of the reflection loss of the specular and diffuse MPCs, and the corresponding fitted models. (a) Specular MPCs. (b) Diffuse MPCs.}\label{fig:rl_modeling}
\end{figure}

\subsection{Material Type}
As discussed in Section~\ref{sec:idenfify}, specular and diffuse MPCs can be separated based on their reflection loss. As illustrated in~\cite{monostatic_lyu,fang_geometry}, material type can be inferred from the reflection loss characteristics. Therefore, in this subsection, we perform a comprehensive analysis and modeling of the reflection loss for both specular and diffuse MPCs reflected from different materials. The MPC clusters associated with different materials are identified by correlating their parameters with the known geometric layout of the measurement scenarios. Note that this analysis focuses exclusively on MPC clusters reflected from a single material type, excluding those that involve reflections from multiple materials.

Fig.~\ref{fig:rl_modeling} presents the CDFs of the reflection loss for both specular and diffuse components, along with their corresponding fitted models. For the specular components, MPCs reflected from metal exhibit the lowest reflection loss, ranging from $0.03$ to $4.25$~dB. Specular reflections from the tile wall show a reflection loss in the range of $[9.41, 10.86]$~dB. The materials glass and cement exhibit comparable reflection loss characteristics, with ranges of $[9.26, 12.50]$~dB and $[10.56, 14.66]$~dB, respectively. Among all the tested materials, polymer demonstrates the highest specular reflection loss, ranging from $11.50$ to $15.58$~dB. As shown in Fig.~\ref{fig:rl_modeling}~(a), the CDFs of specular reflection losses for all materials closely match the fitted normal distributions. Furthermore, the mean reflection losses of specular MPCs for various materials are compared with the material reflection loss database presented in~\cite{fang_geometry}, showing a strong agreement between measured and reference values. Regarding the diffuse components, all diffuse MPCs exhibit higher reflection losses compared to their specular counterparts. The diffuse reflection losses for metal and glass are found to be similar, with ranges of $[15.97,44.90]$~dB and $[15.19,52.53]$~dB, respectively. In contrast, cement, polymer, and tile show comparable diffuse reflection losses, with ranges of $[24.97,65.74]$~dB, $[23.63,63.44]$~dB, and $[29.33,60.99]$~dB, respectively. As illustrated in Fig.~\ref{fig:rl_modeling}~(b), the diffuse reflection loss distributions align well with fitted Weibull models. The distinct statistical characteristics of reflection losses across different materials suggest the feasibility of data-driven material classification. For instance, the models of specular and diffuse reflection loss for various materials can serve as discriminative features for classifiers. A confusion-matrix-based analysis for more materials would be valuable for quantifying classification performance, which we leave for future exploration.

\section{Conclusion}\label{sec:conclusion}
In this work, channel measurements were carried out in three representative indoor scenarios to investigate THz monostatic sensing channels at the $300$~GHz band. The Tx and Rx were co-located to emulate a monostatic sensing configuration, with a total of $57$ TRx positions across all scenarios. A SAGE-based channel parameter estimation algorithm was applied to extract key MPC parameters, including amplitude, delay, and azimuth angle. The resulting channel characteristics were analyzed on a scenario-by-scenario basis. Building on these measurements, we propose the first environment-aware channel modeling framework tailored for monostatic sensing applications. In this framework, key physical attributes, such as reflector quantity, location, geometry, material properties, are systematically mapped to their corresponding channel characteristics. The analysis demonstrates that this framework can reliably infer environmental features from observed channel behavior, validating both its feasibility and accuracy. Future work will extend this research toward AI-assisted, multi-modal, sensing-enhanced channel measurement and modeling, with a focus on ISAC channel modeling in the THz band.

\normalem
\bibliography{monostatic_sensing}

\end{document}